\documentclass[english,tightenlines,eqsecnum,amsmath,amssymb,aps,prd,superscriptaddress,nofootinbib,showpacs,floats]{article}
\usepackage{amsmath,amsfonts,amssymb,multicol}
\usepackage{graphicx,caption,subcaption}
\usepackage{epstopdf}
\usepackage[usenames]{xcolor}
\usepackage{epstopdf}
\definecolor{AHZ}{rgb}{0.0,1,0.0}
\usepackage[colorlinks,linkcolor=AHZ,citecolor=red,bookmarks]{hyperref}
\usepackage{rotating}
\usepackage[mathscr]{eucal}
\usepackage{graphicx}
\usepackage[T1]{fontenc}
\usepackage{babel}
\usepackage{authblk}
\usepackage{epsfig}
\usepackage{color}
\usepackage{amssymb}
\usepackage{fancyhdr}

\def\nn{\nonumber\\}
\newcommand{\f}[2]{\frac{#1}{#2}}
\def\be{\begin{equation}}
\def\ee{\end{equation}}
\def\bea{\begin{eqnarray}}
\def\eea{\end{eqnarray}}

\def\bwt{\begin{widetext}}
	\def\ewt{\end{widetext}}

\usepackage{amsmath}

\usepackage{amssymb}

\begin{document}
	
	\title{Wormholes in Poincar\`{e} gauge theory of gravity}
	\author{Amir Hadi Ziaie\thanks{ah.ziaie@maragheh.ac.ir}}
	\author{Christian Corda\thanks{cordac.galilei@gmail.com}}
	\affil{{\rm Research~Institute~for~Astronomy~and~Astrophysics~of~ Maragha~(RIAAM), University of Maragheh,  P.~O.~Box~55136-553,~Maragheh, Iran}}
	\affil{{\rm International Institute for Applicable Mathematics and Information Sciences, B. M. Birla Science Centre, Adarsh-nagar, Hyderabad 500063 (India) and Istituto Livi, Via Antonio Marini, 9, 59100 Prato (Italy)}}
	\renewcommand\Authands{ and }
	\maketitle
\begin{abstract}
In the present work, we {study} static spherically symmetric solutions representing wormhole configurations in Poincar\`{e} gauge theory ({{\sf PGT}}). The gravitational sector of the Lagrangian is chosen as a subclass of {\sf PGT} Lagrangians for which, the spin-$0^+$ is the only propagating torsion mode. The spacetime torsion in {\sf PGT} has a dynamical nature even in the absence of intrinsic angular momentum (spin) of matter, hence, {torsion can play a principal role in the case of usual spin-less gravitating systems. We therefore} consider a spin-less matter distribution with an anisotropic energy momentum tensor ({\sf EMT}) as the supporting source for wormhole structure {to obtain a class of zero tidal force wormhole solutions.} It is seen that the matter distribution obeys the physical reasonability conditions, i.e., the weak ({\sf WEC}) and null ({\sf NEC}) energy conditions either at the throat and throughout the spacetime. We further consider varying equations of state in radial and tangential directions via definitions $w_r(r)=p_r(r)/\rho(r)$ and $w_t(r)=p_t(r)/\rho(r)$ and {investigate} the behavior of state parameters {at the throat of wormhole}. We observe that our solutions allow for wormhole configurations without the need of exotic matter. Observational features of the wormhole solutions are also discussed utilizing gravitational lensing effects. It is found that the light deflection angle diverges at the throat (which indeed, effectively acts as a photon sphere) and can get zero and negative values depending on the model parameters.
\end{abstract}
\maketitle
\section{Introduction}
One of the most fascinating features of general relativity ({\sf GR}) that has attracted many researchers so far, is the possible existence of hypothetical geometries which have nontrivial topological structure, known as wormholes. {The first attempt toward understanding the concept of wormhole was carried out by Ludwing Flamm in 1916~\cite{Flamm1916} where he noticed that the spatial part of the Schwarzschild spacetime describes something like a bridge or a shortcut between two worlds or two parts of the same world. However, it became clear that the wormhole solutions proposed by him are not traversable and hence are not relativistically applicable. In 1935, a specific wormhole-type solution was proposed by Einstein and Rosen~\cite{ERose}. Their motivation was to construct an elementary particle model represented by a \lq{}\lq{}bridge structure\rq{}\rq{} connecting two identical sheets. This solution was subsequently denoted as Einstein-Rosen bridge.} The concept of {traversable wormholes} was firstly introduced by Misner and Wheeler through their pioneering works~\cite{misner-wheeler,Wheelerworm}, where these objects were obtained from the coupled equations of electromagnetism and {\sf GR} and were denoted {as} \lq{}\lq{}geons\rq{}\rq{}, i.e., gravitational-electromagnetic entities~\cite{misnerwheelerworks}. Wheeler considered Reissner-Nordstrom or Kerr wormholes, as objects of the quantum foam\footnote{Smooth spacetime of {\sf GR} enduring quantum-gravitational fluctuations in topology at the Planck scale.}, which connect different regions of spacetime and operate at the Planck scale. These objects were transformed later into Euclidean wormholes by Hawking~\cite{hawkingwormtrans} and others. However, {the} Wheeler wormholes were properly understood as non-{traversable}~\cite{FulWheel} wormholes\footnote{Since, such type of wormholes do not allow a two way communication between two regions of the spacetime by a minimal surface called the wormhole throat.}, and additionally would, in principle, develop some type of singularity~\cite{GerochJMath}. {The simplest exact traversable wormhole solution was discovered in 1973 by Bronnikov~\cite{Bronn} and independently by Ellis~\cite{Ellisworm}. The wormhole  throat for these solutions is supported by a phantom field which is a scalar field with negative kinetic energy. In a subsequent work, Ellis obtained the evolving version of his solution~\cite{Elliswormdyn}. Also, Clement considered wormhole like solutions in higher dimensional {\sf GR}~\cite{ClementHW}, Einstein-Maxwell theory with an auxiliary scalar field~\cite{EMH} and Einstein-Yang-Mills theory with a ghost Higgs field~\cite{EYMGH}, see also~\cite{FLoboBook,BronnBook} for overviews of wormhole research. However, the interest in wormhole physics was highly stimulated in 1988 after the pioneering work of Morris and Thorne who proposed a new idea of traversable wormhole~\cite{mt,mt1}}. {They} investigated this issue {through} introducing a static spherically symmetric metric and discussed the required conditions for physically meaningful Lorentzian traversable wormholes. However, traversability of a wormhole requires inevitably the violation of null energy condition ({\sf NEC}). In other words, the matter field  providing this geometry is known as exotic matter for which the energy density becomes negative resulting in the violation of {\sf NEC}~\cite{khu}. Although the violation of energy conditions is unacceptable from the common viewpoint of physicists, it has been shown that some effects due to quantum field theory, e.g., Casimir effect can allow for such a violation~\cite{Caseffect}. Also, negative energy densities which are required to support the wormhole configuration may be produced through gravitational squeezing of the vacuum~\cite{negendensqueez}, see also~\cite{khu,Klinkhammer1991} for more details. However, it is generally believed that all classical forms of matter obey the standard energy conditions.
\par
Researches on physics of wormholes by Morris, Thorne, and Yurtsever have opened up, in recent years, a new field of study in theoretical physics and several publications have appeared in recent years among which we can quote, traversable wormhole geometries constructed by matter fields with exotic {\sf EMT}~\cite{MarcoChianese2017}, phantom or quintom-type energy~\cite{phantworm} and wormholes supported by nonminimal interaction between dark matter and dark energy~\cite{intdarksec}, see~\cite{lobocgqreview} for a comprehensive review. {Also an intriguing hypothesis concerns the possibility that the Milky Way could be a huge wormhole~\cite{MilkywR}.} However, owing to the problematic nature of exotic matter, many attempts have been made toward minimizing its usage and {in this way, modified gravity theories~\cite{ExtGravCappo}{, as extensions of {\sf GR},}  can provide a setting in order to overcome} the issue of energy conditions {for} wormhole structures. Much research on wormhole solutions in modified gravity has been done including, wormholes in the framework of gravitational decoupling~\cite{Ortiz2021}, Lovelock theories~\cite{LOVEWORM}, Rastall gravity~\cite{rastallworm},  modified gravities with curvature-matter coupling~\cite{Garcia-Lobo}, scalar-tensor theory~\cite{bd}, $f({\sf R})$ gravity~\cite{fr}, Einstein-Cartan theory ({\sf ECT})~\cite{ectwormhole1,ectwormhole}, Einstein-Gauss-Bonnet~\cite{gmfl} and other theories~\cite{otherworms}.

\par
{An} important issue that motivates one to seek for possible generalizations of {\sf GR} is to provide a correct basis for involving intrinsic angular momentum (spin) of gravitating sources {along with} suitable conservation laws {in} gravitational interactions. As we know, the ingredients of macroscopic matter are elementary particles obeying at least locally, the rules of quantum mechanics and special theory of relativity. As a consequence, all elementary particles can be classified via irreducible unitary representations of the Poincar\`{e} group and can be labeled by mass $m$ and spin $s$. Mass is connected with the translational part of the Poincar\`{e} group and spin with the rotational part. In the microscopic realm of matter, the spin angular momentum becomes important in characterizing the dynamics of matter. One therefore expects that in analogy to coupling of energy-momentum to the spacetime metric, spin is coupled to a fundamental attribute of spacetime and {hence,} plays its own role within the gravitational interactions~\cite{ECT3}. However, in standard {\sf GR}, spin does not couple to any specific geometrical quantity. The simplest generalization of {\sf GR}, in order to incorporate spin contributions is the {\sf ECT} in the context of which the spacetime torsion is physically generated through the presence of spin of matter. As a matter of fact, {in {\sf ECT}} both energy-momentum and spin angular momentum of matter act as sources of the gravitational interaction. However, since the equation governing the torsion tensor is of pure algebraic type, the torsion {field cannot} propagate in the absence of spin effects, namely, outside the matter distribution and thus, in the case of spin-less matter, gravitational equations of {\sf ECT} are identical to those of {\sf GR}~\cite{Venzo}. {In} the past decades, {{\sf PGT}} {has} been developed and {has} become a viable alternative to {\sf GR}. From physical viewpoint (and also geometrically) it is reasonable to consider gravity as a gauge theory of local Poincar\`{e} symmetry of {the} Minkowski spacetime. A formulation of gravity based on local gauge symmetry of spacetime geometry, i.e., the quadratic {\sf PGT}, has been presented in~\cite{ECT3,PGT0,conservation,PGT1,Obukhovgrg,PGT2,PGT3}. In this theory, the gravitational field is described {through an} interacting metric and torsion {field} and is generated by means of {\sf EMT} and spin momentum tensor of gravitating matter. The dynamical feature of spacetime torsion is {determined} by the order of field strength tensors included within the Lagrangian; while the full linear case ({\sf ECT}) bears a non-propagating torsion field, higher order correction terms describe a Lagrangian with dynamical torsion~\cite{PGT1,Obukhovgrg,PGT4}.

\par
Since the advent of {\sf PGT}, isotropic cosmological models have been constructed and studied with the aim of resolving fundamental cosmological problems~\cite{PGTCOSMOLOGICAL}. It is shown that, {imposing certain restrictions on model parameters}, gravitational interaction in the framework of homogeneous isotropic models is altered in comparison {to} {\sf GR} and can be repulsive under certain conditions. {This allows preventing} initial singularity of the Universe as well as explaining the current accelerated expansion of the Universe without resorting to {a dark energy source}. Static spherically symmetric electro-vacuum solutions in {\sf PGT} have been {investigated} in~\cite{PGTelectrovacuum} where it is shown that the spacetime torsion is induced by both the mass and charge of the source. Black hole solutions with dynamical massless torsion in {\sf PGT} have been reported in~\cite{Cembranos2017}. The obtained solutions are of Reissner-Nordstrom type with a Coulomb-like curvature provided by the torsion field, see also~\cite{Obukhovgrg} and references therein. Also, lensing features of a black hole in the framework of {\sf PGT} {have} been studied in~\cite{Akhshabi2021} where it is shown that the presence of spacetime  torsion modifies the deflection angle. In the present work, motivated by the above considerations, we are interested in finding static spherically symmetric solutions representing wormhole geometries in {\sf PGT}. {In section}~\ref{FES}, {we introduce} the field equations of {\sf PGT}. {In section~\ref{WHS1} we} obtain exact wormhole solutions, with zero tidal force, {for an spin-less anisotropic matter distribution} satisfying {\sf WEC} and {\sf NEC}. In Section~\ref{OBSFEATURE} we discuss observational features of the obtained solutions and finally we conclude our paper and point out future works in section~\ref{concluding}.
\section{Field equations for spin-$0^{+}$ mode}\label{FES}
In the present section we give a brief review on {\sf PGT} using the procedure already done within this area~\cite{ECT3,PGT2,PGT3,PGT5,PGT6,PGT7,PGT8,PGT9,F.W.HEHL1987}. The {\sf PGT} is founded on a spacetime with a Riemann-Cartan geometry, i.e., a Lorentz signature metric with a metric compatible connection. According to the ten independent parameters of the Poincar\`{e} group, we have ten gauge potentials. The gravitational field is then described by means of two sets of local gauge potentials, {the four of which are} the translation group i.e., the tetrad field ${\sf e}_{i}^{\,\,\mu}$ and the metric compatible connection $\Gamma_{i\mu}^{\,\,\,\,\,\nu}$ which is associated with the six gauge potentials of the Lorentz group. The corresponding field strengths are the spacetime torsion
\bea\label{SPTORCUR}
{\sf Q}_{ij}^{\,\,\,\,\,\mu}&=&2\left(\partial_{[i\,\,\,\,j]}\!\!\!\!\!\!{\sf e}^{\,\,\,\,\mu}+\Gamma_{[i|\nu\,\,\,\,\,\,\,|j]}\!\!\!\!\!\!\!\!{\sf e}^{\!\!\!\!\!\!\!\mu\,\,\,\,\,\,\,\,\,\,\nu}\right),
\eea
for the tetrads and the spacetime curvature
\bea\label{SPTORCUR1}
{\sf R}_{ij\mu}^{\,\,\,\,\,\,\,\,\nu}&=&2\left(\partial_{[i\,\,\,\,j]}\!\!\!\!\!\!\Gamma^{\,\,\,\,\,\nu}_{\,\,\mu}+\Gamma_{[i|\sigma\,\,\,\,\,\,\,|j]}\!\!\!\!\!\!\!\!\!\Gamma_{\,\,\,\,\,\,\mu} ^{\!\!\!\!\!\!\!\!\nu\,\,\,\,\,\,\,\,\,\,\,\,\,\sigma}\right),
\eea
for the connection. These quantities obey the Bianchi identities
\be\label{Bianchii}
\nabla_{[i\,\,\,\,\,\,\,\,\,\,\,\,\,]}\!\!\!\!\!\!\!\!\!\!\!{\sf Q}_{jk}^{\,\,\,\,\,\,\,\,\mu}\equiv{\sf R}_{[ijk]}^{\,\,\,\,\,\,\,\,\,\,\mu},~~~~~~~~\nabla_{[i\,\,\,\,\,\,\,\,\,\,\,\,\,]}\!\!\!\!\!\!\!\!\!\!\!{\sf R}_{jk}^{\,\,\,\,\,\,\,\,\mu\nu}\equiv0,
\ee
where $\nabla_i$ is the covariant derivative associated to the connection $\Gamma_{i\mu}^{\,\,\,\,\,\nu}$. The Greek indices denote local Lorentz indices and the Latin ones are coordinate indices. The tetrads satisfy the following equalities
\bea\label{tetradeq}
{\sf e}^{i}_{\,\,\mu}{\sf e}_{i}^{\,\,\nu}=\delta_{\mu}^{\,\,\nu},~~~~{\sf e}^{i}_{\,\,\mu}{\sf e}_{j}^{\,\,\mu}=\delta_{j}^{i},~~~~{\sf g}_{ij}={\sf e}_{i}^{\,\,\mu}{\sf e}_{j}^{\,\,\nu}\eta_{\mu\nu}.
\eea
The conventional action which is invariant under the Poincar\`{e} gauge group can be put into the {following} form 
\be\label{actionpgt}
{\sf A}=\int d^4x{\sf e}\left({\mathcal L}_G+{\mathcal L}_M\right),
\ee
where ${\mathcal L}_M={\mathcal L}_M\left({\sf e},\Gamma,\Psi,\partial\Psi\right)$ stands for the minimally coupled Lagrangian density of matter fields ($\Psi$) which determines the energy momentum and spin source currents, ${\mathcal L}_G={\mathcal L}_G\left({\sf e}_{i}^{\,\,\mu},\partial_j{\sf e}_{i}^{\,\,\mu},\Gamma_{i\mu}^{\,\,\,\,\,\nu},\partial_j\Gamma_{i\mu}^{\,\,\,\,\,\nu}\right)={\mathcal L}_G\left({\sf e}_{i}^{\,\,\mu},{\sf Q}_{ij}^{\,\,\,\,\mu},{\sf R}_{ij}^{\,\,\,\,\mu\nu}\right)$ being the gravitational Lagrangian density and ${\sf e}={\rm det}\left({\sf e}_{i}^{\,\,\mu}\right)$. As demonstrated within the aforementioned works, the field equations can be derived from the action (\ref{actionpgt}) by performing independent variations with respect to the gauge potentials. These equations can then be written as
\bea
\nabla_j{\sf H}_{\mu}^{\,\,\,ij}-{\sf E}_\mu^{\,\,\,i}&=&{\sf T}_{\!\mu}^{\,\,i},\label{FESPGT}\\
\nabla_j{\sf P}_{\mu\nu}^{\,\,\,\,\,\,ij}-{\sf U}_{\mu\nu}^{\,\,\,\,\,\,i}&=&{\sf S}_{\mu\nu}^{\,\,\,\,\,\,i},\label{FESPGT1}
\eea
with the field momenta 
\bea
{\sf H}_{\mu}^{\,\,\,ij}&:=&\f{\partial{\sf e}{\mathcal L}_G}{\partial\partial_j{\sf e}_i^{\,\,\mu}}=2\f{\partial{\sf e}{\mathcal L}_G}{\partial{\sf Q}_{ji}^{\,\,\,\,\mu}},\label{momenta1}\\
{\sf P}_{\mu\nu}^{\,\,\,\,\,\,ij}&:=&\f{\partial{\sf e}{\mathcal L}_G}{\partial\partial_j\Gamma_i^{\,\,\mu\nu}}=2\f{\partial{\sf e}{\mathcal L}_G}{\partial{\sf R}_{ji}^{\,\,\,\,\mu\nu}},\label{momenta2}
\eea
and
\bea\label{momenta3}
{\sf E}_\mu^{\,\,\,i}:={\sf e}^i_{\,\mu}{\sf e}{\mathcal L}_G-{\sf Q}_{\mu j}^{\,\,\,\,\,\,\nu}{\sf H}_{\nu}^{\,\,\,ji}-{\sf R}_{\mu j}^{\,\,\,\,\,\,\nu\sigma}{\sf P}_{\nu\sigma}^{\,\,\,\,\,\,ji},~~~~~~~{\sf U}_{\mu\nu}^{\,\,\,\,\,\,i}:={\sf H}_{[\nu\mu]}^{\,\,\,\,\,\,\,\,\,i}.
\eea
Variation of the matter Lagrangian leaves us with the following expressions for the source terms
\be\label{source}
{\sf T}_{\!\mu}^{\,\,i}=\f{\partial{\sf e}{\mathcal L}_{m}}{\partial{\sf e}_i^{\,\mu}},~~~~~~~{\sf S}_{\mu\nu}^{\,\,\,\,\,\,i}=\f{\partial{\sf e}{\mathcal L}_{m}}{\partial\Gamma_i^{\,\mu\nu}},
\ee
which are known, respectively as the Noether energy-momentum and spin density currents. As a consequence of minimal coupling principle, these two tenors satisfy suitable energy-momentum and angular momentum conservation laws~\cite{conservation}. As usual, the Lagrangian is assumed to be, at most, quadratic in the field strengths. Therefore, the field momenta can be expressed by linear combinations of the field strengths as
\bea
{\sf H}_{\mu}^{\,\,\,ij}&=&\f{{\sf e}}{\ell^2}\sum_{n=1}^{3}a_n \overset{\!\!\!\!\!\!\!\!\!(n)}{{\sf Q}^{\,ji}_{\,\,\,\,\,\mu}},\\
{\sf P}_{\mu\nu}^{\,\,\,\,\,\,ij}&=& -\f{a_0{\sf e}}{\ell^2}{\sf e}^i_{\,\,[\mu\,\,\,\,\,\,\,\,\,\,}\!\!\!\!\!\!\!\!{\sf e}^j_{\,\nu]}+\f{{\sf e}}{\kappa}\sum_{n=1}^{6}b_n\overset{\!\!\!\!\!\!\!\!\!\!(n)}{{\sf R}^{ji}_{\,\,\,\,\mu\nu}},
\eea
where $\overset{\!\!\!\!\!\!\!\!\!(n)}{{\sf Q}^{\,ji}_{\,\,\,\,\,\mu}}$ and $\overset{\!\!\!\!\!\!\!\!\!\!(n)}{{\sf R}^{ji}_{\,\,\,\,\mu\nu}}$ are the algebraically irreducible parts of the torsion and curvature tensors, respectively; $\ell$ and $\kappa$ are coupling constants and $a_n$, $b_n$ are free coupling parameters. The torsion tensor is decomposed into its three irreducible components known as the vector, axial and tensor {parts} as
\be\label{torsionirreducible}
{\sf Q}_i={\sf Q}_{ij}^{\,\,\,\,j},~~~~{\sf Z}_i=\f{1}{2}\epsilon_{ijkm}{\sf Q}^{jkm},~~~~{\sf D}_{ijk}={\sf Q}_{i(jk)}-\f{1}{3}{\sf Q}_i{\sf g}_{jk}+\f{1}{3}{\sf g}_{i(j\,\,\,\,\,\,\,\,\,\,)}\!\!\!\!\!\!\!\!\!\!{\sf Q}_{k}~,
\ee
whence the torsion tensor can be re-expressed as
\be\label{torsionrex}
{\sf Q}_{ijk}=\f{4}{3}{\sf D}_{[ij]k}+\f{2}{3}{\sf Q}_{[i\,\,\,\,\,\,\,\,]}\!\!\!\!\!\!\!{\sf g}_{j\,\,k}+\f{1}{3}\epsilon_{ijkm}{\sf Z}^m.
\ee
In {\sf PGT}, in addition to the dynamical nature of spacetime metric (described by the translational gauge potential) the rotational gauge potential assumes some independent dynamics. The various dynamic modes in {\sf PGT}, beyond those of metric, were first studied through the linearized theory~\cite{PGT7,PGT10}. The dynamics of connection, which can be described by the torsion tensor, is represented in terms of six modes with certain spins and parity as, $2^\pm$, $1^\pm$ and $0^\pm$. A reasonable dynamic mode should transport positive energy and should not propagate outside the forward null cone. {This criterion is} often referred to as absence of ghost and tachyon. Investigations of the linearized quadratic {\sf PGT} revealed that at most three modes can be simultaneously dynamic. The results of Hamiltonian analysis also found to be consistent with those of linearized investigation~\cite{PGT3,HamiltonianPGT}. A careful scrutiny of the Hamiltonian and propagation {modes}~\cite{spin0+,PGT11,PGT12,PGT13} led to conclusion that the effects due to nonlinearities could be expected to render all of these cases physically unacceptable, with the exception of two scalar connection modes with spin-$0^+$ and spin-$0^-$. In this regard, a cosmological model (with flat {\sf FLRW} spacetime) has been studied in~\cite{PGT14,Shie2008} and it was found that the $0^+$ mode naturally couples to the acceleration of the Universe and could account for current observations. The extension of this model to include spin-$0^-$ mode has also been studied in~\cite{spin0-}, see also~\cite{Baekler2011} for a beautiful generalization of torsion cosmology models. In the present study we only consider the simple spin-$0^+$ case. {We then} choose $a_2=-2a_1$, $a_3=-a_1/2$ and except for $b_6\neq0$, we assume {that} all $b_n$ coefficients {are zero}, see also~\cite{spin0+} for more details. The associated gravitational Lagrangian density for this mode then reads
\bea\label{ecfieldeq}
{\mathcal L}_G=-\f{a_0}{2}{\sf R}+\f{b_6}{24}{\sf R}^2+\f{a_1}{8}\left[{\sf Q}_{\nu\sigma\mu}{\sf Q}^{\nu\sigma\mu}+2{\sf Q}_{\nu\sigma\mu}{\sf Q}^{\mu\sigma\nu}-4{\sf Q}_{\mu}{\sf Q }^{\mu}\right],~~~~~{\sf Q}_\mu={\sf Q}_{\mu\nu}^{\,\,\,\,\,\,\,\nu},
\eea
where physical reasonability on kinetic energy requires that $a_1>0$ and $b_6>0$. Moreover, the Newtonian limit requires $a_0=-(8\pi G)^{-1}=-1$~\cite{PGT7}, where we have set the unites so that $8\pi G=c=1$. For a vanishing spin source (${\sf S}_{\mu\nu}^{\,\,\,\,\,\,i}=0$) one can perform variation of gravitational Lagrangian (\ref{ecfieldeq}) with respect to the gauge potentials. This gives, for Eq. (\ref{FESPGT1}), the following equations~\cite{Shie2008}
\bea\label{fesspin0+}
\nabla_{\nu}{\sf R}=-\f{2}{3}\left[{\sf R}+\f{6\mu}{b_6}\right]{\sf Q}_\nu,~~~~~{\sf Z}_\nu=0,~~~~~~{\sf D}_{\mu\nu\sigma}=0,
\eea
where $\mu=a_1-a_0$ is the effective mass of the linearized $0^+$ mode. The second and third parts of Eq. (\ref{fesspin0+}) leave us with the following constraint on torsion tensor
\be\label{torsionconstraint}
{\sf Q}_{ij}^{\,\,\,\,\,\mu}=\f{2}{3}{\sf Q}_{[i\,\,\,\,\,\,\,\,]}\!\!\!\!\!\!\!{\sf e}_{j}^{\,\,\,\,\,\mu},
\ee
with the help of which, Eq. (\ref{FESPGT}) can be rewritten as~\cite{Shie2008}
\bea\label{HEQreex}
\nabla_j{\sf H}_\mu^{\,\,\,ij}-{\sf E}_{\mu}^{\,\,i}={\sf e}\Bigg\{\!\f{2a_1}{3}\left[{\sf e}^i_{\,\nu}\nabla_\mu{\sf Q}^\nu-{\sf e}^i_{\,\mu}\tilde{\nabla}_j{\sf Q}^j\right]\!+\!{\sf e}^i_{\,\mu}\left[\f{a_0}{2}{\sf R}-\f{b_6}{24}{\sf R}^2+\f{a_1}{3}{\sf Q}_i{\sf Q}^i\right]\!+\!{\sf R}_\mu^{\,\,i}\left(\f{b_6}{6}{\sf R}-a_0\right)\!\!\!\Bigg\}\!\!=\!\!{\sf T}_\mu^{\,\,i}.
\eea
Next, in order to better deal with Eq. (\ref{HEQreex}) and first part of (\ref{fesspin0+}), one can rewrite them in terms of metric ${\sf g}_{jk}$ and torsion ${\sf Q}_{ij}^{\,\,\,k}$. By doing so, {we arrive} at the following field equations~\cite{Shie2008}
\bea
&&a_0\tilde{{\sf G}}_{ij}+\bar{{\sf T}}_{ij}=-{\sf T}_{ij},\label{FESFinal}\\
&&\tilde{\nabla}_i{\sf R}+\f{2}{3}\left[{\sf R}+\f{6\mu}{b_6}\right]{\sf Q}_i=0,\label{FESFinal1}
\eea
where $\tilde{\nabla}_i$ stands for covariant derivative with respect to Levi-Civita connection $\tilde{\Gamma}_{ij}^{\,\,\,\,k}$ and $\tilde{{\sf G}}_{ij}$ is the standard Einstein tensor. The tensor $\bar{{\sf T}}_{ij}$ represents contribution due to {the} scalar torsion mode and is given by
\be\label{contscalartorsion}
\bar{{\sf T}}_{ij}=-\f{\mu}{3}\left[\tilde{\nabla}_i{\sf Q}_j+\tilde{\nabla}_j{\sf Q}_i-2{\sf g}_{ij}\tilde{\nabla}_k{\sf Q}^k\right]-\f{\mu}{9}\left[2{\sf Q}_i{\sf Q}_j+{\sf g}_{ij}{\sf Q}_k{\sf Q}^k\right]-\f{b_6}{6}{\sf R}\left[{\sf R}_{(ij)}-\f{1}{4}{\sf g}_{ij}{\sf R}\right],
\ee
where, {the} Ricci curvature tensor and Ricci scalar {are given by}, respectively
\bea\label{RICCITS}
{\sf R}_{ij}&=&\tilde{{\sf R}}_{ij}+\f{1}{3}\left[2\tilde{\nabla}_j{\sf Q}_i+{\sf g}_{ij}\tilde{\nabla}_k{\sf Q}^k\right]+\f{2}{9}\left[{\sf Q}_i{\sf Q}_j-{\sf g}_{ij}{\sf Q}_k{\sf Q}^k\right],\\
{\sf R}&=&\tilde{{\sf R}}+2\tilde{\nabla}_i{\sf Q}^i-\f{2}{3}{\sf Q}_i{\sf Q}^i.
\eea

\section{Wormhole Solutions}\label{WHS1}
Let us consider the general static and spherically symmetric line element representing a wormhole spacetime given by
\begin{eqnarray}\label{evw}
ds^2=-{\rm e}^{2\Phi(r)}dt^2+\left(1-\f{b(r)}{r}\right)^{-1}dr^2+r^2d\Omega^2,
\end{eqnarray}
where $d\Omega^2=d\theta^2+\sin^2\theta d\phi^2$ is the standard line element on a unit two-sphere, $\Phi(r)$ is the redshift function and $b(r)$ is the wormhole shape function. The radial coordinate ranges from $r_0$ (wormhole\rq{}s throat) to spatial infinity. At the throat, defined by the condition $r_0=b(r_0)$, there is a coordinate singularity where the radial metric component ${\sf g}_{rr}$ diverges, however, the radial proper distance
\be\label{radpropdis}
\ell(r)=\pm\int_{r_0}^{r}\f{dr}{\left(1-b(r)/r\right)^{1/2}},
\ee
is required to be finite. Indeed, at the throat we have $\ell(r_0)=0$ while, $\ell<0 (>0)$ on the left (right) side of the throat. Conditions on redshift and shape functions under which, wormholes are traversable have been discussed completely in~\cite{mt}. Traversability of the wormhole requires that the spacetime be free of horizons which are defined as the surfaces with ${\rm e }^{2\Phi(r)}\rightarrow0$; therefore the redshift function must be finite everywhere. In the present work, we try to find $b(r)$, assuming there is no tidal force present, i.e., $\Phi(r)={\sf Constant}$ and we will set this constant to be zero for latter convenience. The non-vanishing components of the torsion tensor are given as~\cite{Cembranos2017}
\be\label{tornonvcs}
{\sf Q}_{tr}^{\,\,\,\,\,r}={\sf Q}_{t\theta}^{\,\,\,\,\,\theta}={\sf Q}_{t\phi}^{\,\,\,\,\,\phi}=\f{B(r)}{3},~~~~{\sf Q}_{tr}^{\,\,\,\,\,t}={\sf Q}_{\theta r}^{\,\,\,\,\,\theta}={\sf Q}_{\phi r}^{\,\,\,\,\,\phi}=\f{B^{\prime}(r)}{2B(r)},
\ee
where {a prime denotes $d/dr$. We} note that these components satisfy the constraints given in the second and third parts of Eq. (\ref{fesspin0+}). Let us define the time-like and space-like vector fields, respectively as $u^i=[1,0,0,0]$ and $v^i=\left[0,\sqrt{1-b(r)/r},0,0\right]$, so that $u^iu_i=-1$ and $v^jv_j=1$. The anisotropic {\sf EMT} of matter source then takes the form
\be\label{emtaniso}
{\sf T}_{ij}=[\rho(r)+p_t(r)]u_iu_j+p_t(r){\sf g}_{ij}+[p_r(r)-p_t(r)]v_iv_j,
\ee
with $\rho(r)$, $p_r(r)$, and $p_t(r)$ being the energy density, radial and tangential pressures, respectively. Before proceeding further, it is worth mentioning that there are a number of ways to construct the wormhole structure. One method is a purely geometric approach in which the metric components are considered as pre-determined functions
in order to {obtain} the desired wormhole geometry. Therefore, the supporting {\sf EMT} is completely specified from the geometry through the field equations. Such an strategy for constructing wormhole configuration has been employed in~\cite{mt}. In the present study we follow this strategy and write the components of the field equation (\ref{FESFinal}) as 
\bea
\rho(r)&=&\f{b_6(r-b)^2B^{\prime\prime2}}{8r^2B^2}-\Bigg\{\f{b_6(r-b)^2B^{\prime2}}{4r^2B^3}+\f{b_6(r-b)(rb^\prime-4r+3b)B^\prime}{8r^3B^2}\nn
&+&\f{(r-b)\left(3b_6b^\prime+r^2(b_6B^2+9\mu)\right)}{9r^3B}\Bigg\}B^{\prime\prime}+\f{3b_6(r-b)^2B^{\prime4}}{32r^2B^4}+\f{b_6(r-b)\left(rb^\prime-4r+3b\right)B^{\prime3}}{8r^3B^3}\nn&+&\Bigg\{b_6(r-b)r^3B^2+{27}b_6b^2-72(\mu r^2+b_6)rb+24(3\mu r^2+2b_6)r^2-6b_6rbb^\prime+3b_6r^2b^{\prime2}\Bigg\}\f{B^{\prime2}}{96r^4B^2}\nn
&+&\f{9b_6b^{\prime2}-6r^2\left(9a_0-b_6B^2\right)b^\prime+r^4B^2(b_6B^2+18\mu)}{54r^4}\nn&+&\f{\left[3b_6b^\prime+r^2(b_6B^2+9\mu)\right](rb^\prime-4r+3b)}{18r^4B}B^\prime,\label{rhofeq}
\eea
\bea
p_r(r)&=&\f{3b_6(r-b)^2B^{\prime\prime2}}{8r^2B^2}+\f{9b_6(r-b)^2B^{\prime4}}{32r^2B^4}+\f{b_6(r-b)\left[3rb^\prime-{8}r+{5}b\right]B^{\prime3}}{8r^3B^3}-\Bigg\{\f{3b_6(r-b)^2B^{\prime2}}{4r^2B^3}\nn&-&\f{b_6(r-b)\left[3rb^\prime-4r+b\right]B^{\prime}}{8r^3B^2}+\f{b_6(r-b)\left[rb^\prime-b+\f{2}{9}r^3B^2\right]}{2r^4B}\Bigg\}B^{\prime\prime}\nn&+&\Bigg\{27b_6r^2b^{\prime2}+18b_6r(4r-7b)b^\prime+40b_6r^3(r-b)B^2-117b_6b^2\nn&-&216r\left[\mu r^2-\f{4}{3}b_6\right]b+216r^2\left[\mu r^2-\f{2}{3}b_6\right]\Bigg\}\f{B^{\prime2}}{288r^4B^2}\nn&+&\Bigg\{9b_6r^2b^{\prime2}+2b_6r\left[r^3B^2-6r-3b\right]b^\prime-2b_6r^3bB^2\nn&-&27b_6b^2+\left(18b_6r-36\mu r^3\right)b+36\mu r^4\Bigg\}\f{B^\prime}{36r^5B}\nn&+&\f{27b_6rb^{\prime2}+12b_6\left[r^3B^2-\f{9}{2}b\right]b^\prime+\left[b_6r^3B^4-18(\mu r^3+b_6b)B^2+162a_0b\right]r^2}{162r^5},\label{prfeq}
\eea
\bea
p_t(r)&=&-\f{b_6(r-b)^2B^{\prime\prime2}}{8r^2B^2}-\f{3b_6(r-b)^2B^{\prime4}}{32r^2B^4}-\f{b_6(r-b)\left[rb^\prime-2r+b\right]}{8r^3B^3}B^{\prime3}+\Bigg\{\f{b_6(r-b)^2B^{\prime2}}{4r^2B^3}\nn&+&\f{b_6(r-b)(rb^\prime-b)B^\prime}{8r^3B^2}+\f{(r-b)\left[12\mu r^3+b_6rb^\prime-3b_6b\right]}{12r^4B}\Bigg\}B^{\prime\prime}\nn&-&\Big\{b_6r^2b^{\prime2}+b_6(4r-6b)rb^\prime+\f{8}{9}b_6r^3(r-b)B^2-11b_6b^2\nn&-&4(6\mu r^2-7b_6)rb+8(3\mu r^2-2b_6)r^2\Big\}\f{B^{\prime2}}{32r^4B^2}\nn&-&\Bigg\{b_6r^2b^{\prime2}+\left[\left(12\mu r^2+4b_6\right)r^2-8b_6rb\right]b^{\prime}\nn&+&\f{8b_6}{3}r^3(r-b)B^2-9b_6b^2+12(\mu r^2+b_6)rb-24\mu r^4\Bigg\}\f{B^{\prime}}{24r^5B}\nn&+&\f{\left[3b_6r^3B^2+81a_0r^3+27b_6b\right]b^\prime+\left[b_6r^3B^4+9(b_6b-2\mu r^3)B^2-81a_0b\right]r^2}{162r^5}.\label{ptfeq}\nn
\eea
We note that the metric functions must be determined so that the field equation (\ref{FESFinal1}) be satisfied. To this aim we write the temporal and radial components of this equation as
\bea
&&\left[1-\f{b}{r}\right]\left({B^{\prime2}-2BB^{\prime\prime}}\right)+\left[\f{b^\prime}{r}-\f{4}{r}+\f{3b}{r^2}\right]BB^\prime+\f{4}{9}\left[\f{3b^\prime}{r^2}+B^2+\f{9\mu}{b_6}\right]B^2=0,\label{rtcomps7}\\
&&\left[9\left(1-\f{b}{r}\right)\f{B^\prime}{B^2}+\f{9rb^\prime-12r+3b}{2r^2B}\right]B^{\prime\prime}-\f{9(r-b)B^{\prime3}}{2rB^3}-\f{9rb^\prime-24r+15b}{2r^2B^2}B^{\prime2}\nn&+&\Big\{9b_6r^2b^{\prime\prime}+\left[4b_6r^3B^2-36\mu r^3+{6}b_6rb^\prime+36b_6r-54b_6b\right]\Big\}\f{B^\prime}{6b_6r^3B}\nn&-&\f{3r^2(r-b)B^{\prime\prime\prime}-2rBb^{\prime\prime}+4b^{\prime}B}{r^3B}=0,\label{rtcomps71}
\eea
{The above set of differential equations} can be solved simultaneously for the shape function with a general solution given by
\bea\label{gensolb}
b(r)=\left\{ {C}_1-\f{1}{3}\int r F_1(r) {\rm e}^{3\!\int \!{F_2}(r) dr}dr \right\}{\rm e}^{-3\!\int \!{F_2}(r)dr},
\eea
where
\bea\label{FF12}
F_1(r)&=&\f{4rB^3+\f{36\mu r}{b_6}B-18rB^{\prime\prime}+9r\f{B^{\prime2}}{B}-36B^\prime}{3rB^\prime+4B},\nn
F_2(r)&=&\f{2rBB^{\prime\prime}-rB^{\prime2}+3BB^{\prime}}{B(3rB^{\prime}+4B)},
\eea
and $C_1$ is an integration constant. Now, assuming a power-law behavior for {the} torsion, $B(r)=B_0r^n$ ($n<0$ and $n\neq-4/3,-2$), the integration can be performed giving
\bea\label{bfinalsol}
b(r)=\f{3n(n+2)}{3n^2+6n+4}r-\f{4\mu}{b_6(n+2)^2}r^3-\f{4B_0^2}{3(9n^2+20n+12)}r^q+C_1r^m,
\eea
{where $q=2n+3$ and $m=-\f{3n(n+1)}{3n+4}$.} The spherical surface $r=r_0$ have to satisfy the following fundamental conditions: $b(r_0)=r_0$, $b(r)<r$ for $r>r_0$ and $rb^\prime-b<0$ which is known as the flare-out condition~\cite{mt,mt1}. The second condition guarantees that the (Lorentzian) metric signature is preserved for $r>r_0$. In order to find the integration constant we use the condition $b(r_0)=r_0$ at the wormhole throat.  We therefore get the integration constant as
\be\label{intc}
C_1=\f{4B_0^2}{3(9n^2+20n+12)}r_0^{\f{9n^2+20n+12}{3n+4}}+\f{4\mu}{b_6(n+2)^2}r_0^{\f{3(n+2)^2}{3n+4}}+\f{4}{3n^2+6n+4}r_0^{\f{3n^2+6n+4}{3n+4}}.
\ee
Also, {the} flare-out condition leads to the following inequality at the throat 
\bea\label{flarecond}
b^\prime(r)\Big|_{r=r_0}\!\!\!\!\!\!<1\implies\frac{4r_0\left[b_6\left({B_0}^2 {r_0}^{2 n+2}+3\right)+9 \mu {r_0}^2\right]}{3 {b_6} (3n+4)}>0.
\eea
\par
\begin{figure}
\begin{center}
\includegraphics[scale=0.25]{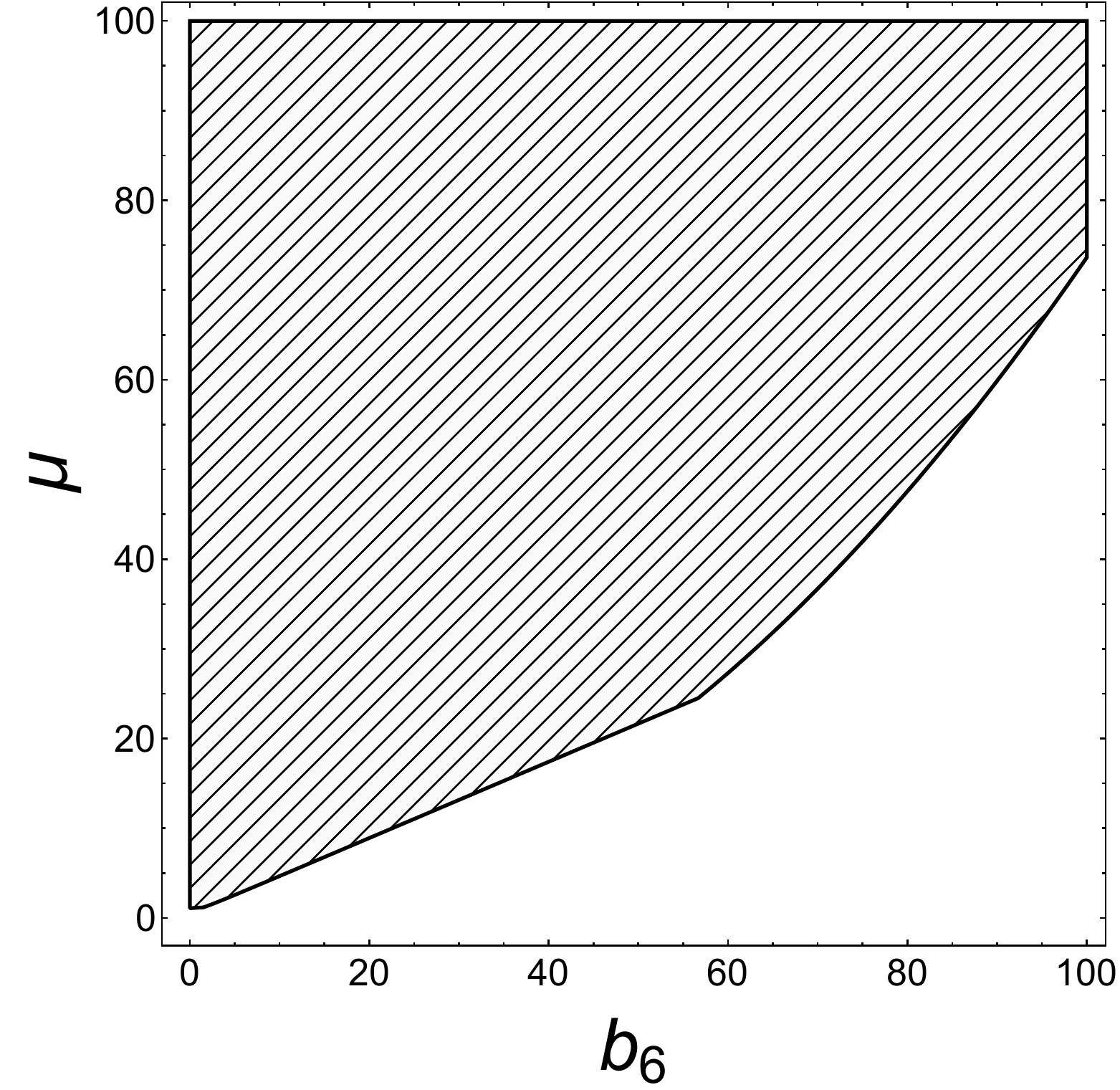}
\includegraphics[scale=0.26]{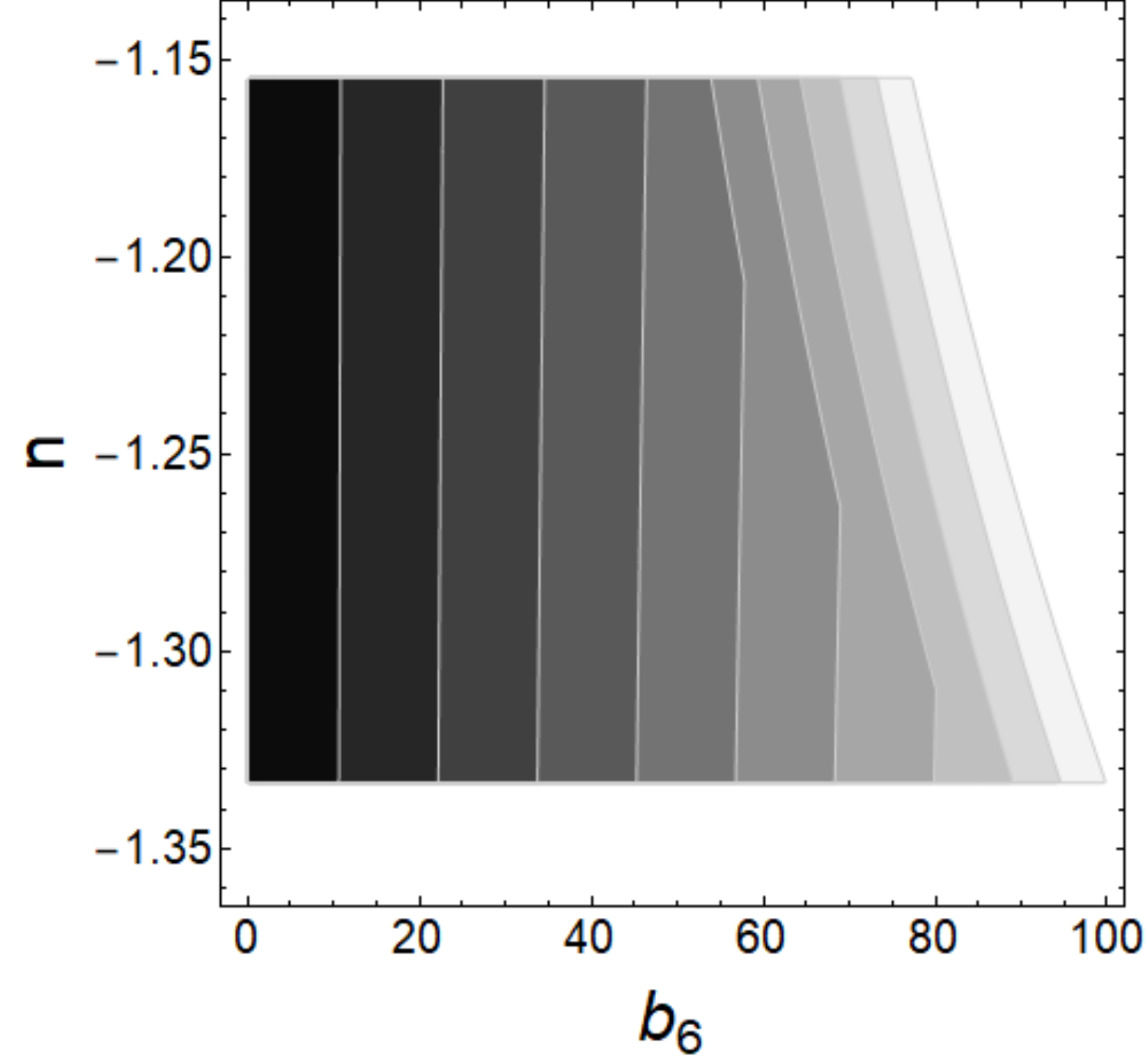}
\hspace*{8cm}\includegraphics[scale=0.26]{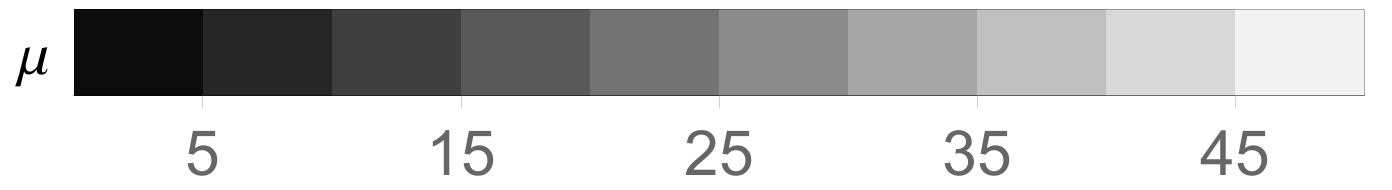}
\caption{Left panel: The allowed values of $\mu$ and $b_6$ parameters (shaded region) for $B_0=0.1$, $n=-1.20$, $r_0=1$ and $a_0=-1$. Right panel: The allowed values of $n$ and $b_6$ parameters for different values $\mu$ parameter.}\label{fig1}
\end{center}
\end{figure}

Next we proceed to obtain the energy density, radial and tangential pressures for our solution. These quantities take the form 
\bea\label{energysol}
\rho(r)=\beta_1-(2n+3)(a_0+\mu)\alpha_3r^{2n}-(a_0+\mu)\alpha_1r^{-2}+\f{3n(n+1)(a_0+\mu)}{3n+4}\alpha_4r^{\f{-3(n+2)^2}{3n+4}}
\eea
\bea\label{prrsol}
p_r(r)=\gamma_1+(a_0+\mu)\alpha_3r^{2n}+(a_0+\mu)\alpha_1r^{-2}+(a_0+\mu)\alpha_4r^{\f{-3(n+2)^2}{3n+4}},
\eea
\bea\label{prtsol}
p_t(r)=\gamma_1+(n+1)(a_0+\mu)\alpha_3r^{2n}-\f{(a_0+\mu)(3n^2+6n+4)}{2(3n+4)}\alpha_4r^{\f{-3(n+2)^2}{3n+4}},
\eea
where
\bea\label{betas}
\beta_1&=&\f{3\mu (8a_0-\mu(n^2+4n-4))}{2b_6(n+2)^2},~~~~~~~~~~\alpha_1=\f{3n(n+2)}{3n^2+6n+4}\\
\gamma_1&=&\f{\mu (-8a_0+\mu(3n^2+12n+4))}{2b_6(n+2)^2},~~~~~~\alpha_2=-\f{4\mu}{b_6(n+2)^2},\\
\alpha_3&=&-\f{4B_0^2}{3(9n^2+20n+12)}.
\eea
In the framework of classical {\sf GR}, the fundamental flaring-out condition results in violation of {\sf NEC}. Such a violation can be surveyed by applying the focusing theorem on a congruence of null rays, defined by a null vector field $k^\mu$, where $k^\mu k_\mu=0$~\cite{khu,FLoboBook}. For the {\sf EMT} given in (\ref{emtaniso}) the {\sf NEC}
is given by 
\bea
&&\rho(r)+p_{r}(r)\geq0,~~~~~\rho(r)+p_{t}(r)\geq0\label{nec}.
\eea
{Also,} for the sake of physical reliability of the solutions, we require that the wormhole configuration respects the {\sf WEC} given by the following inequalities
\bea
&&\rho(r)\geq0,~~~~~\rho(r)+p_{r}(r)\geq0,~~~~~\rho(r)+p_{t}(r)\geq0.\label{wec}
\eea
Using expressions (\ref{energysol})-(\ref{prtsol}) we then get
\bea
\rho(r)+p_{r}(r)\!\!\!\!&=&\!\!\!\!\beta_1+\gamma_1-2(n+1)(a_0+\mu)\alpha_3r^{2n}+\f{3n^2+6n+4}{3n+4}(a_0+\mu)\alpha_4r^{\f{-3(n+2)^2}{3n+4}},\label{rhopluspr}\\
\rho(r)+p_{t}(r)\!\!\!\!&=&\!\!\!\!\beta_1+\gamma_1-(n+2)(a_0+\mu)\alpha_3r^{2n}-(a_0+\mu)\alpha_1r^{-2}+\f{(a_0+\mu)(3n^2-4)}{2(3n+4)}\alpha_4r^{\f{-3(n+2)^2}{3n+4}}.\label{rhoplusp}\nn
\eea
Thus, the energy conditions at the throat take the form
\bea
\!\!\!\!\rho(r)\Big|_{r=r_0}\!\!\!\!\!\!\!\!\!\!\!\!&=&\!\!\!\!\frac{8 {B_0}^2 {b_6} ({a_0}+\mu ) {r_0}^{2 n+2}+9 \left[-2 {a_0} {b_6} n+8 {a_0} \mu  {r_0}^2-2 {b_6} \mu  n+\mu ^2 (4-3 n) {r_0}^2\right]}{6 {b_6} (3 n+4) {r_0}^2}\geq0,\label{energy0}\\
\!\!\!\!\rho(r)+p_{r}(r)\Big|_{r=r_0}\!\!\!\!\!\!\!\!\!\!\!\!&=&\!\!\!\!\frac{4 ({a_0}+\mu ) \left[{b_6} \left({B_0}^2 {r_0}^{2 n+2}+3\right)+9 \mu  {r_0}^2\right]}{3 {b_6} (3 n+4){r_0}^2}\geq0,\label{rhopluspr0}\\
\!\!\!\!\rho(r)+p_{t}(r)\Big|_{r=r_0}\!\!\!\!\!\!\!\!\!\!\!\!&=&\!\!\!\!\frac{({a_0}+\mu ) \left[{b_6} \left(2 {B_0}^2 {r_0}^{2 n+2}-9 n-6\right)+18 \mu {r_0}^2\right]}{3 {b_6} (3 n+4) {r_0}^2}\geq0\label{rhopluspt0}.
\eea
The set of parameters ${\sf M}=\left\{n,b_6,\mu,B_0\right\}$ construct a 4-dimensional parameter space that the allowed regions of which are determined through physically reasonable conditions on wormhole configuration. We therefore require that: {{i})} the shape function satisfies the flare-out condition throughout the spacetime and also at the throat, i.e., the inequality (\ref{flarecond}) holds; ${ii)}$ the spacetime be free of horizons; ${iii)}$ the supporting matter for wormhole configuration obeys the {\sf WEC}, i.e., inequalities given in (\ref{wec}) must be satisfied at the throat and throughout the spacetime. We note that {\sf WEC} implies the null form. Hence, in order that the energy conditions be satisfied at the throat the inequalities (\ref{energy0})-(\ref{rhopluspt0}) must be fulfilled. Moreover, the coefficients of $r$ within the expressions (\ref{energysol}), (\ref{rhopluspr}) and (\ref{rhoplusp}) must be positive so that the energy conditions hold throughout the spacetime. We then proceed to find possible bounds on the space parameter {\sf M} subject to the above mentioned conditions. As discussed in~\cite{Shie2008}, there is not too much constraint on the $\mu$ parameter, except for its positivity and finiteness as a mass parameter\footnote{{In the linearized {\sf PGT}, the $\mu$ parameter is recognized as the mass of one of the propagating torsion modes with zero spin and positive parity that is, an ordinary scalar, which has many attractive features~\cite{Hehl1204,Shirafuji1435}.}}, since the baryonic matter will only interact with the scalar torsion mode indirectly by gravitation. We also demand $n<0$ so that the torsion converges asymptotically. The left panel in Fig. (\ref{fig1}) presents a 2D subspace of the 4D parameter space constructed out of the allowed values of $\mu$ and $b_6$ parameters. For any point within the shaded region of this subset of {\sf M} the conditions ({i})-({iii}) are respected. In the right panel, another subset of {\sf M}, which satisfies conditions ({i})-({iii}), has been sketched in terms of the pair of parameters $(n,b_6)$. Each region corresponds to a specific value of $\mu$ parameter (as shown in the bar legend) so that, the larger the value of this parameter, the greater the allowed area for $(n,b_6)$ parameters.
\par
In order to estimate the asymptotic behavior of radial metric component, i.e., ${\sf g}_{rr}=\left(1-b(r)/r\right)^{-1}$ we note that as $r\rightarrow\infty$, the last two terms within the solution (\ref{bfinalsol}) vanish for {$q<0$ and $m<0$. This requires $-4/3<n<-1$}.  {We then get 
\be\label{brrAss}
1-\f{b(r)}{r}\Bigg|_{r\rightarrow\infty}=\f{4}{\zeta}\left[1+\f{\mu\zeta}{b_6(n+2)^2}r^2\right],~~~\zeta=3n^2+6n+4,~~~\zeta>0.
\ee
Hence, asymptotically,} the radial component of the metric mimics a de-Sitter-like or anti de-Sitter-like {behavior} depending on the signs of $\mu$ and $b_6$ parameters. In Hamiltonian analysis of scalar modes of {\sf PGT}~\cite{spin0+}, the positivity of kinetic energy requires that $b_6>0$ and $a_1>0$. The last condition also {demands} $\mu>0$. Thus, {asymptotically, $g_{rr}$ shows an anti de-Sitter-like behavior. A de-Sitter-like behavior can also occur} if one relaxes the condition on positivity of kinetic energy allowing thus for $\mu<0$. This case {can be of interest in} cosmological scenarios with negative kinetic energy~\cite{cosscenariosLambda}. {It should be noted that the asymptotic flatness of the wormhole spacetime depends on mass parameter of the scalar torsion mode. As can be seen from expression (\ref{brrAss}), the spacetime is asymptotically non-flat if $\mu\neq0$, i.e., the case for which the spin-$0^+$ propagating mode is massive~\cite{Shirafuji1435}. In the present work, for $-4/3<n<-1$, the case $\mu=0$ is the only possible case of asymptotically flat spacetime for which the propagating torsion mode is massless. It may also be possible to find other exact asymptotically flat solutions, e.g., considering a pre-determined function for the wormhole shape, that satisfies flare-out and asymptotic flatness conditions, and then solve the resulting differential equations (\ref{rtcomps7}) and (\ref{rtcomps71}) in order to find the functionality of the torsion field; or, seeking for asymptotic flat solutions for the shape function by taking other suitable forms for the torsion function $B(r)$. However, due to the complexity of the field equations, this issue remains as an open problem and can be the subject of further studies. We also note that the vanishing of $\mu$ parameter does not necessarily mean that the wormhole configuration has no effect on particle trajectories around it as the shape function is nonzero in the region $r_0<r<\infty$.}
\begin{figure}
\begin{center}
\includegraphics[scale=0.20]{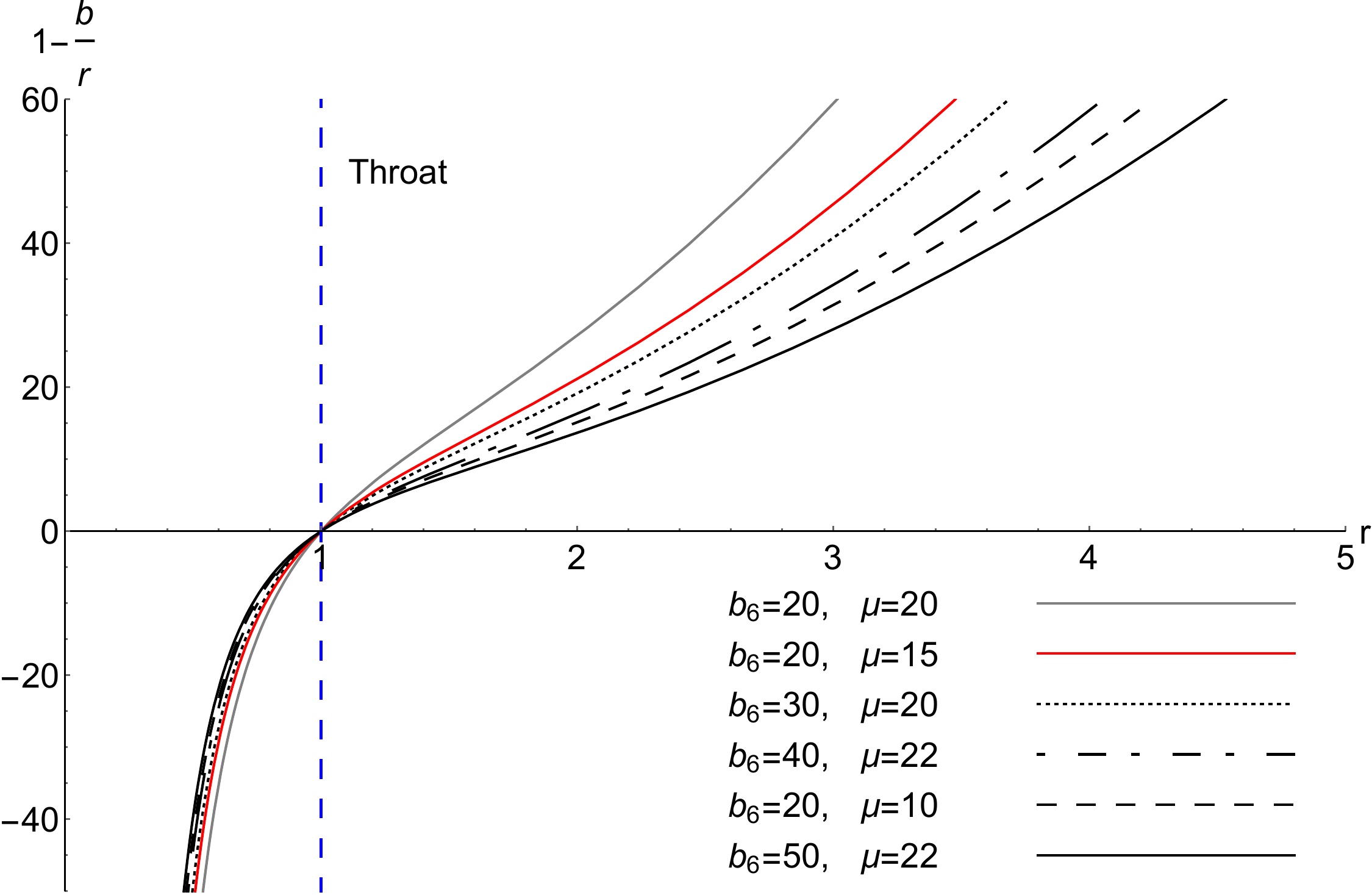}
\includegraphics[scale=0.20]{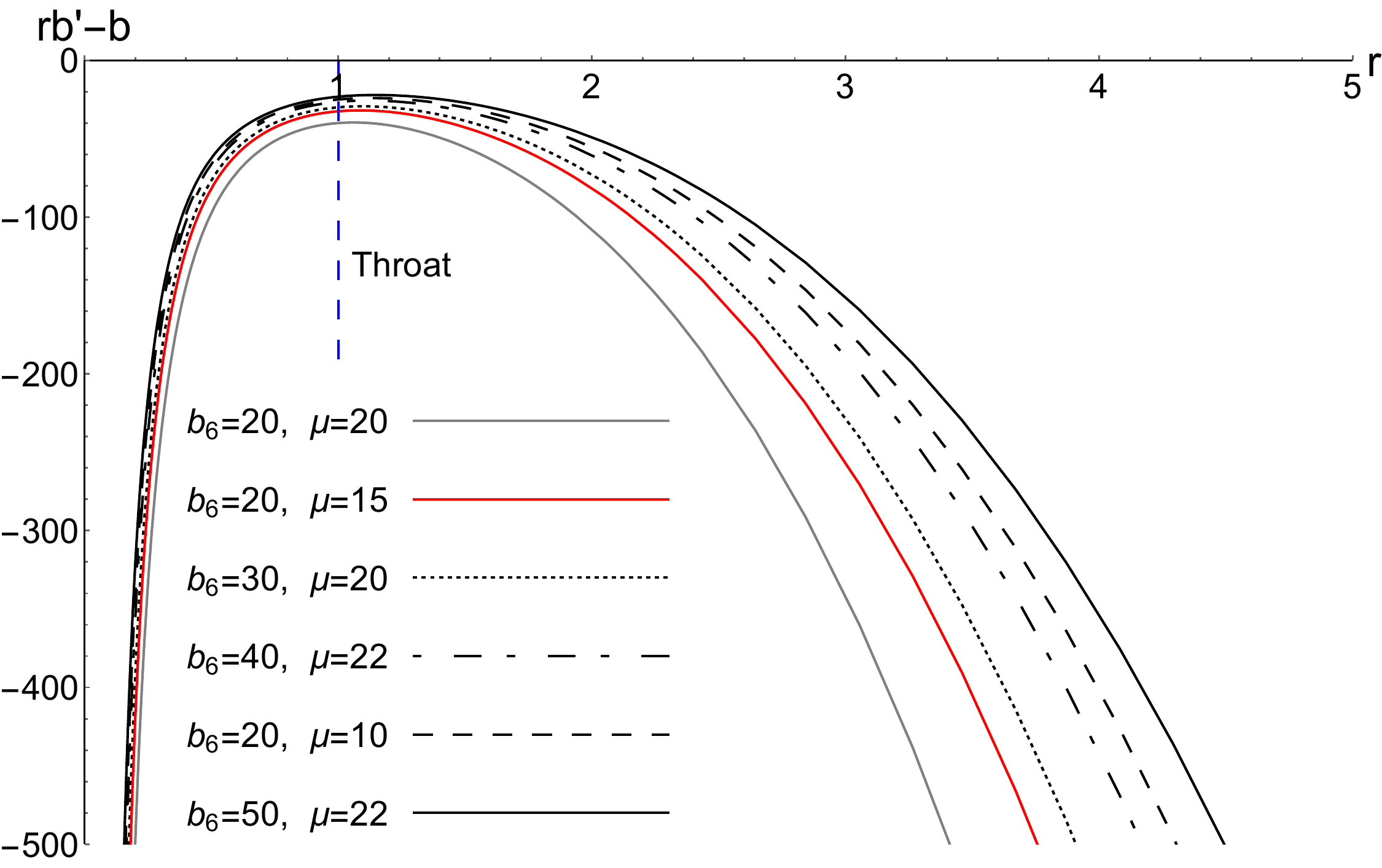}
\caption{Left panel: Behavior of inverse of radial metric component for $n=-1.2$ and different values of $(b_6,\mu)$ parameters. The blue dashed line indicates the location of the throat. Right panel: Plot of flare-out condition for $n=-1.2$ and different values of the parameters $b_6$ and $\mu$. We have set $B_0=0.1$, $r_0=1$ and $a_0=-1$.}\label{fig2}
\end{center}
\end{figure}
\par
\begin{figure}
\begin{center}
\includegraphics[scale=0.22]{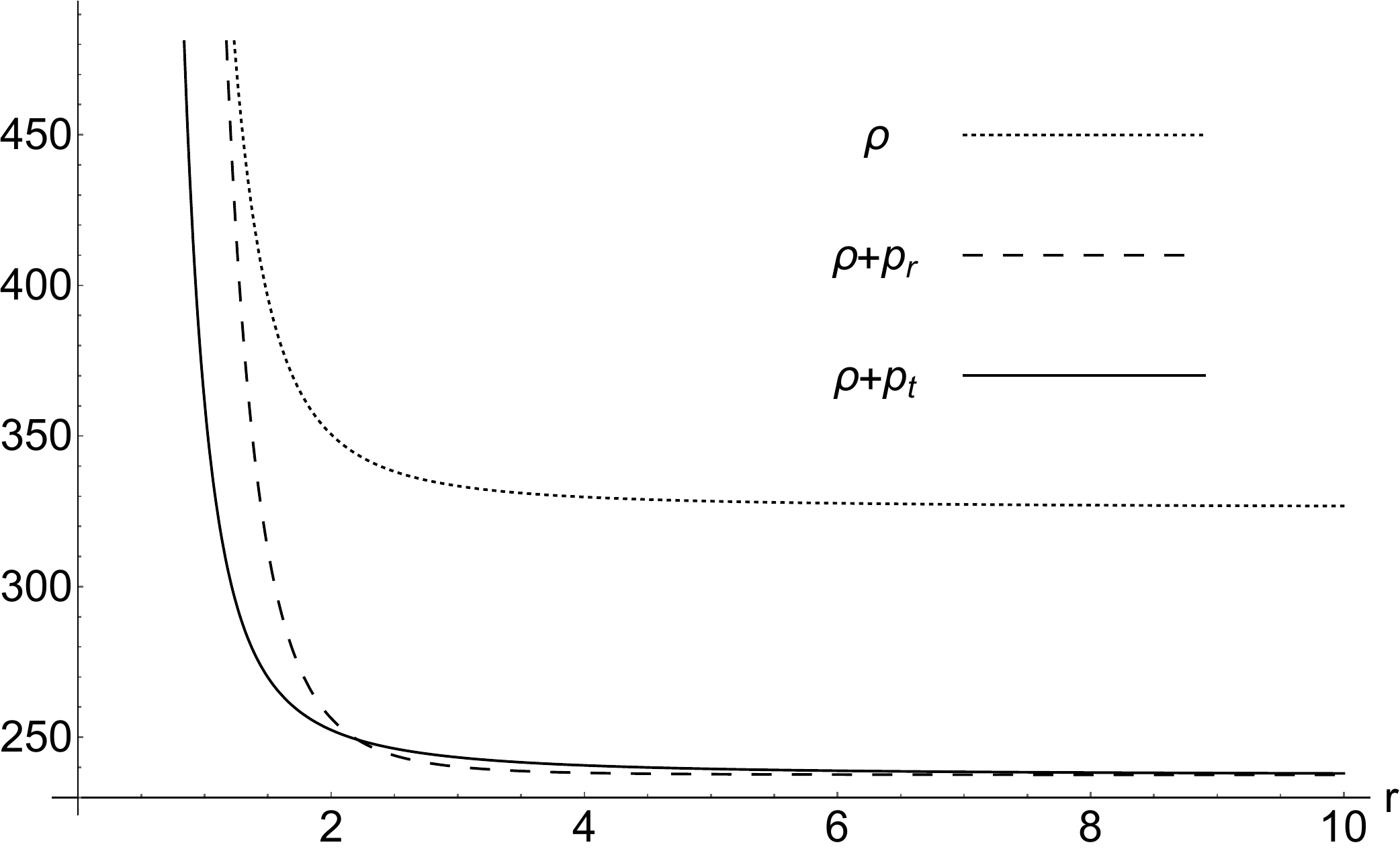}
\includegraphics[scale=0.22]{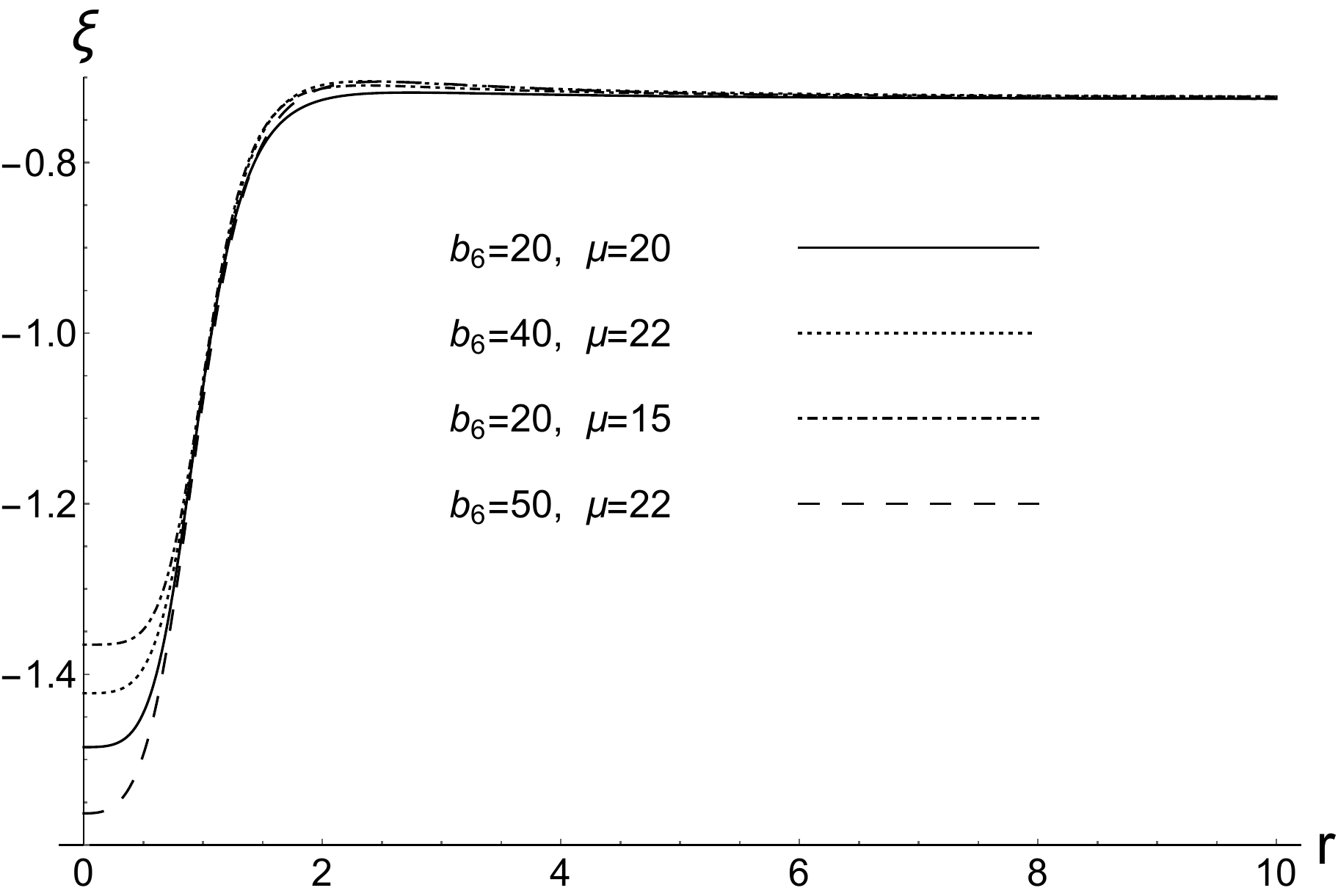}
\caption{Left panel: Plot of {\sf WEC} for $n=-1.2$ and $b_6=\mu=20$. Right panel: Plot of exoticity parameter against radial coordinate for $n=-1.2$ and different values of $(b_6,\mu)$ parameters. We have set $B_0=0.1$, $r_0=1$, $a_0=-1$.}\label{fig3}
\end{center}
\end{figure}

The Left panel of Fig. (\ref{fig2}) presents the inverse of radial metric component (${\sf g}_{rr}^{-1}$), {where} we observe that ${\sf g}_{rr}^{-1}$ is positive for $r>r_0$ and thus the metric signature is preserved {in this range}. To be a {wormhole} solution, one needs to impose that the throat flares out. This can be seen at the right panel of Fig. (\ref{fig2}) as the satisfaction of flare-out condition. The left panel in Fig. (\ref{fig3}) shows that the energy density and quantities $\rho+p_r$ and $\rho+p_t$ remain positive for the allowed values of Fig. (\ref{fig1}); thus the {\sf WEC} and {\sf NEC} are satisfied throughout the spacetime. Following the results of \cite{mt}, one may define a measure of exoticity of matter through the exoticity parameter $\xi$, given as~\cite{mt},\cite{FLoboBook}
\be\label{exoticity}
\xi(r)=\f{\tau(r)-\rho(r)}{\mid\!\!\rho(r)\!\!\mid}=-\f{\rho(r)+p_r(r)}{\mid\!\!\rho(r)\!\!\mid},
\ee
where, $\tau(r)=-p_r(r)$ is the radial tension. The positiveness of $\xi(r)$ signals exotic behavior of the matter. The right panel in Fig. (\ref{fig3}) shows the behavior of exoticity parameter for allowed values of model parameters. It is seen that this parameter is negative at the throat and stays negative for $r>r_0$. Thus, there is no need of introducing exotic matter in order to construct the present wormhole solutions.

\begin{figure}
\begin{center}
\includegraphics[scale=0.18]{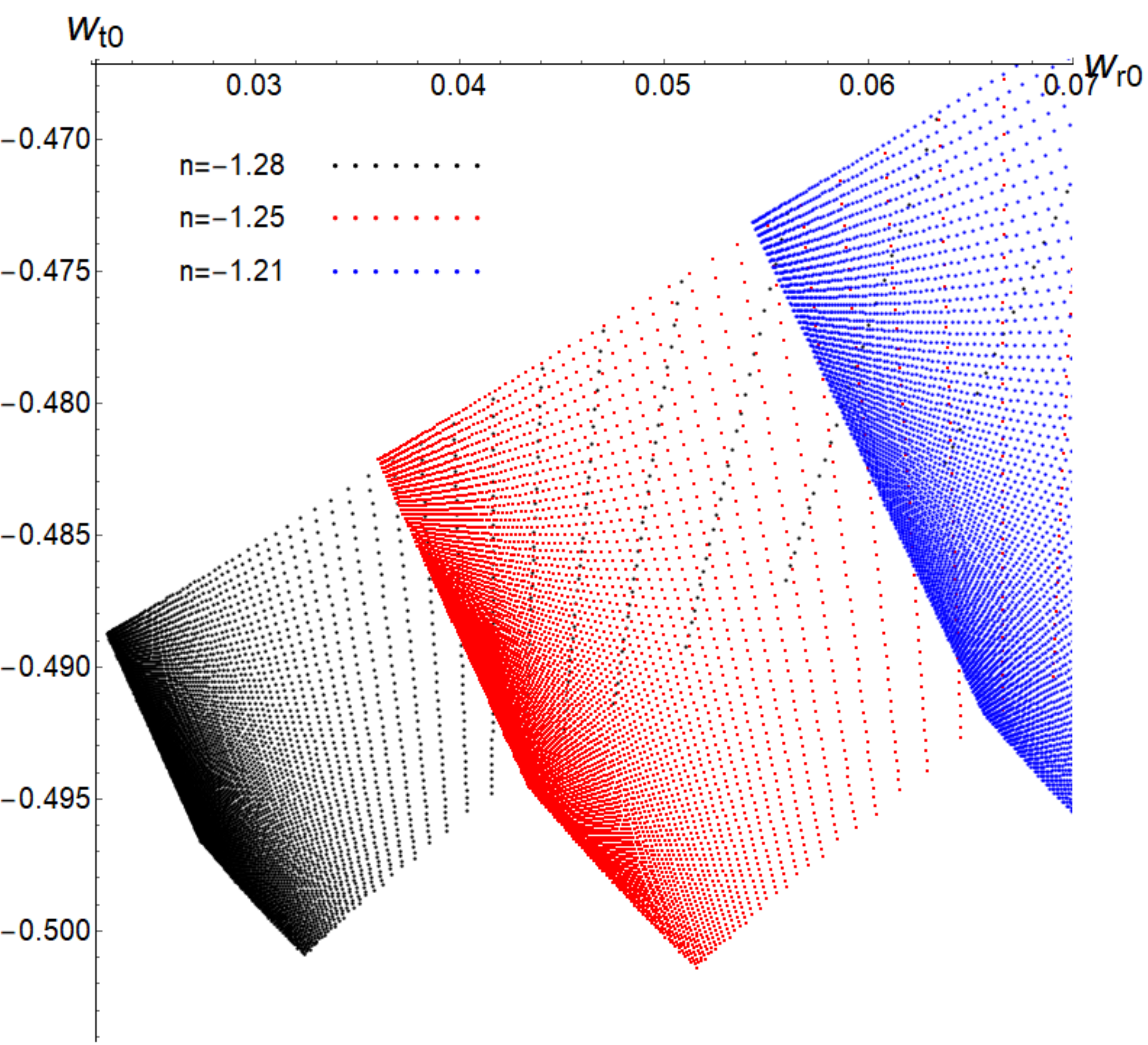}
\includegraphics[scale=0.22]{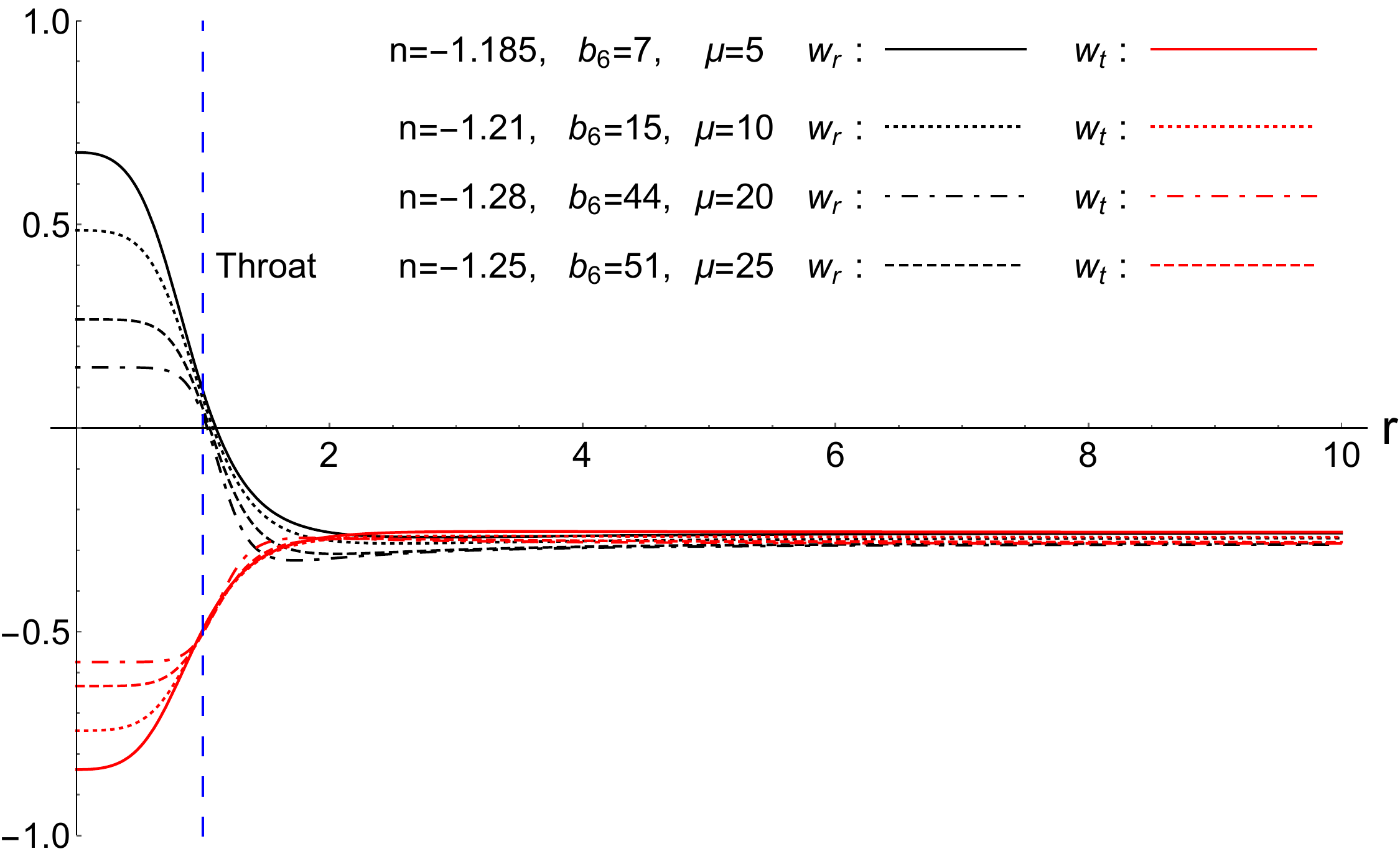}
\caption{Left panel: The allowed values of state parameters at the wormhole throat for different values of $n$ parameter. Right panel: The behavior of state parameters $w_r(r)$ (black curves) and $w_t(r)$ (red curves) for different values of model parameters. The blue dashed line indicates the location of the throat. We have set $B_0=0.1$, $r_0=1$ and $a_0=-1$. }\label{fig4}
\end{center}
\end{figure}

\begin{figure}
\begin{center}
\includegraphics[scale=0.24]{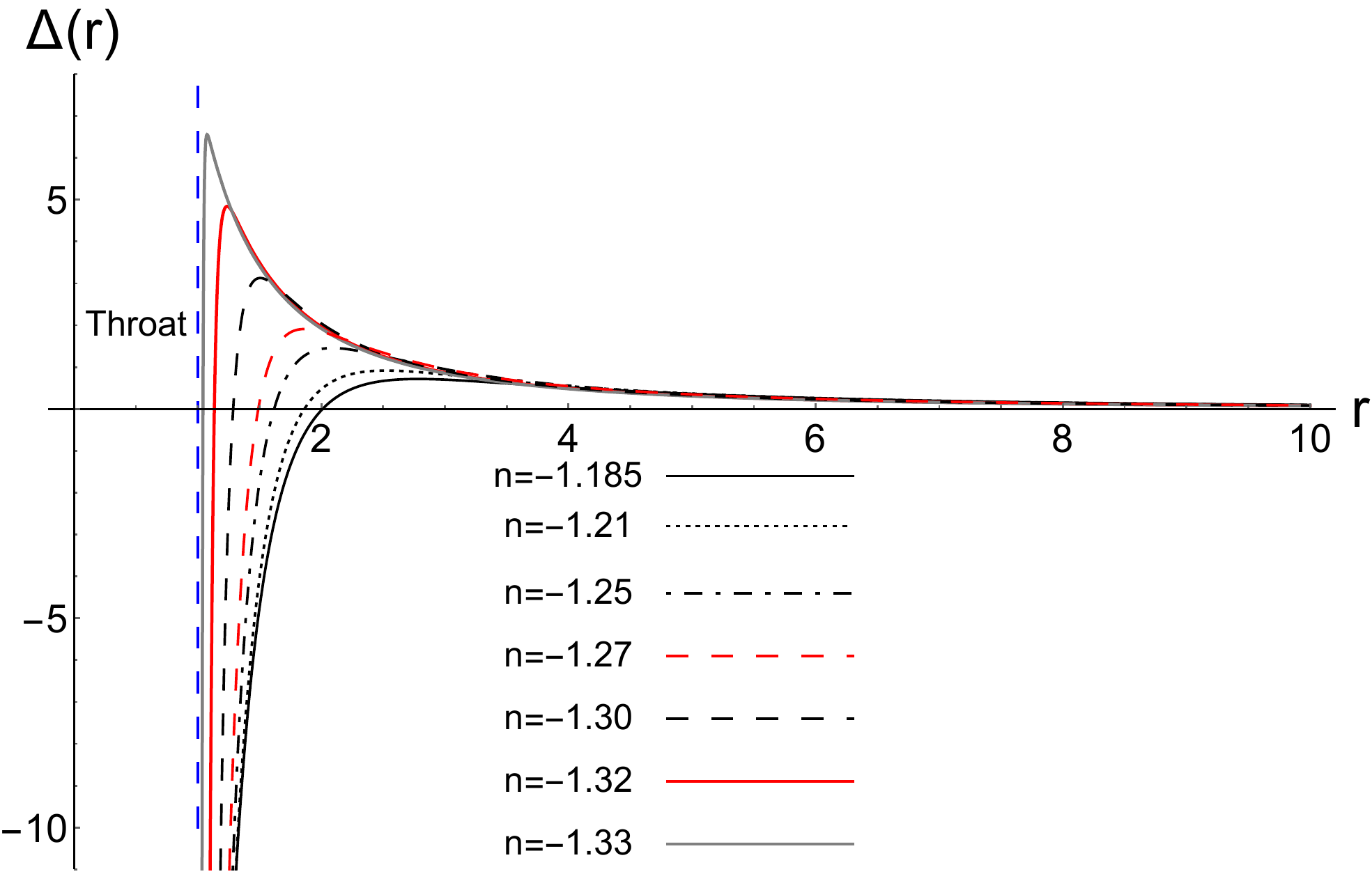}
\includegraphics[scale=0.24]{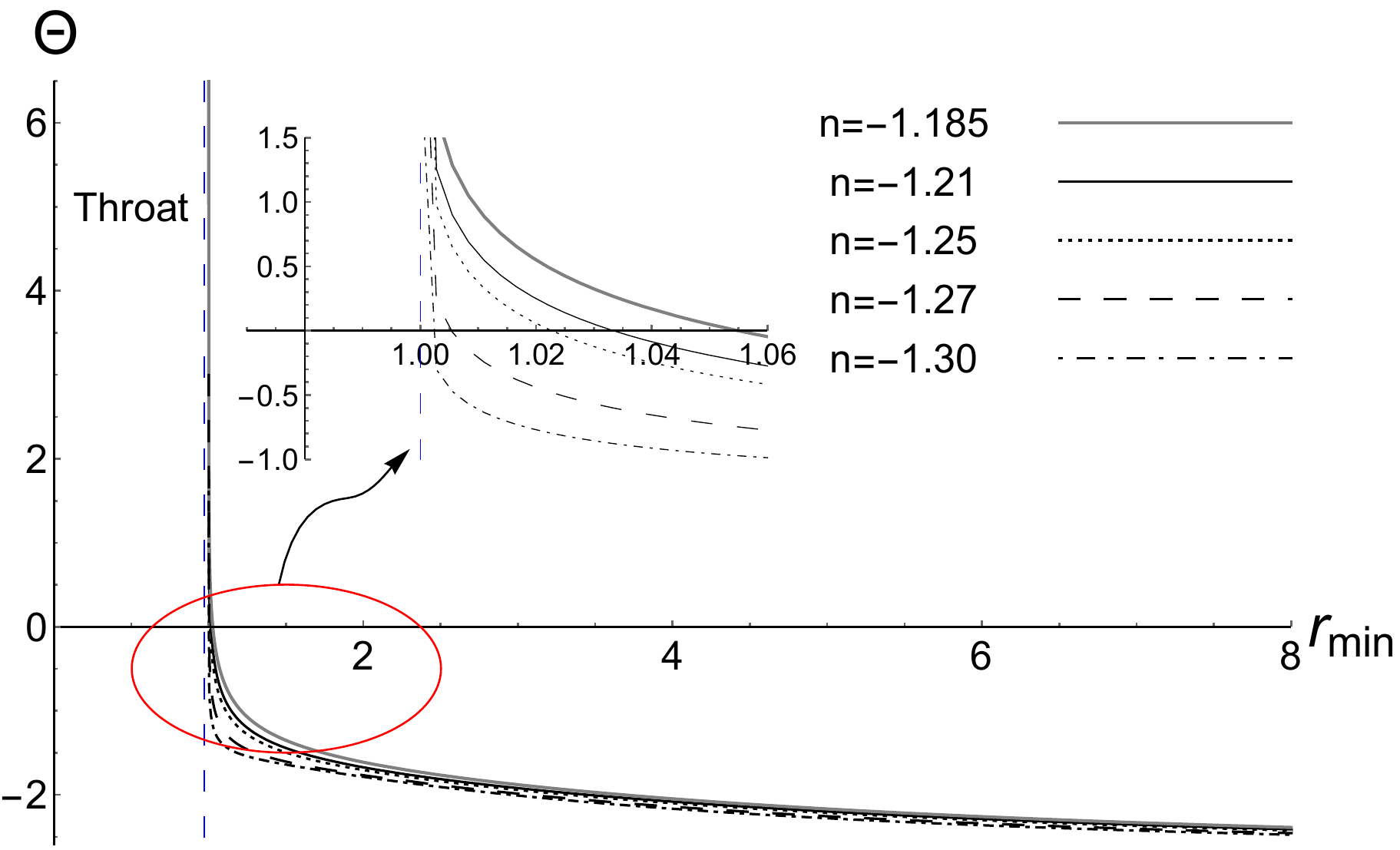}
\caption{Left panel: The behavior of anisotropy parameter against the radial coordinate for different values of parameter $n$. Right panel: Deflection angle against closest distance approach for different values of $n$ parameter. We have set $B_0=0.1$, $b_6=10$, $\mu=5$, $r_0=1$ and $a_0=-1$. The blue dashed line indicates the location of the throat.}\label{fig5}
\end{center}
\end{figure}

The wormhole configurations we have presented so far respect {\sf WEC} and {\sf NEC} with a non-exotic fluid as the supporting matter. In order to get a better understanding of the behavior of such a type of matter we proceed with considering a linear relation between pressure profiles and energy density with $r$-dependent state parameters as
\be\label{eoss}
p_r(r)=w_r(r)\rho(r),~~~~~~p_t(r)=w_t(r)\rho(r).
\ee
The state parameters will take the following form at the throat
\bea\label{wr0wt0}
w_{r0}=w_r(r_0)&=&\frac{3 (3 n+4) \left[2 {a_0} {b_6}+\mu  \left(2 {b_6}+3 \mu  r_0^2\right)\right]}{8 B_0^2 {b_6} ({a_0}+\mu ) r_0^{2 n+2}-9 \left[2 {a_0} \left({b_6} n-4 \mu  r_0^2\right)+\mu  \left(2 {b_6} n+\mu  (3 n-4) r_0^2\right)\right]},\\
w_{t0}=w_t(r_0)&=&\frac{-4 B_0^2 {b_6} ({a_0}+\mu ) r_0^{2 n+2}-3 \left[4 {a_0} \left({b_6}+3 \mu  r_0^2\right)+\mu  \left(4 b_6-9 \mu  n r_0^2\right)\right]}{8 B_0^2 b_6 (a_0+\mu ) r_0^{2 n+2}+9 \left(-2 a_0 b_6 n+8 a_0 \mu  r_0^2-2 b_6 \mu  n+\mu ^2 (4-3 n) r_0^2\right)}.
\eea
At first glance, depending on the model parameters, the radial and tangential state parameters at wormhole throat can assume different values. However, these values have to fulfill the conditions on physical validity of the wormhole solution, i.e., conditions ({i})-({iii}). To this aim, we proceed with finding the allowed values for the pair $(b_6,\mu)$ for different values {of} $n$ parameter (see the left panel in Fig. (\ref{fig1})). Then, we are able to {evaluate} the values of state parameters at the throat subject to the conditions stated above. By doing so, we obtain the bounds on state parameters at the throat as shown in the left panel of Fig. (\ref{fig4}). We therefore observe that the state parameters obey {the ranges} $0.022\lessapprox w_{r0}\lessapprox0.07$ and $-0.5\lessapprox w_{t0}\lessapprox-0.46$, respectively, hence the wormhole supporting matter respects the {\sf WEC} and {\sf NEC} at the throat. {The right panel shows the behavior of state parameters versus radial coordinate where, we observe that the wormhole spacetime tends to an isotropic configuration in the limit $r\rightarrow\infty$}. This behavior {can be also verified} through the {anisotropy }parameter, defined as, $\Delta(r)=p_t(r)-p_r(r)=\left[w_t(r)-w_r(r)\right]\rho(r)$. Since $\rho(r)>0$ throughout the spacetime, it is the sign of the term in square brackets that {decides} the geometry of wormhole configuration. Let us define the coordinate radius $r_2$ so that for $r_0<r_2<\infty$ we have $\Delta(r_2)=0$, see the left panel of Fig.~(\ref{fig5}). We therefore note that since the ratio $2\Delta/r$ represents the force due to anisotropic nature of the configuration, we have an attractive geometry for $r_0<r<r_2$ and a repulsive geometry for $r_2<r<\infty$. The value of coordinate radius $r_2$ depends on model parameters, specifically for the present case, the smaller the absolute value of $n$ parameter, the larger the value of coordinate radius $r_2$. Moreover, having passed the negative values ($r>r_2$), the anisotropy parameter reaches a maximum {value at $r=r_3$ where $\Delta^\prime(r_3)=0$}. As the left panel shows, the larger the absolute value of $n$ parameter, the closer the maximum value of anisotropy to the throat. One then may intuitively imagine that the rate of growth of spacetime torsion around the throat could affect the anisotropy of the wormhole configuration. As $r\rightarrow\infty$ we have $\Delta(r)\rightarrow0$, thus asymptotically, the supporting matter of the wormhole configuration tends to an isotropic fluid.   

\section{Gravitational Lensing Effects}\label{OBSFEATURE}
One of the interesting ways for detecting wormholes is to search for their gravitational lensing effects. In the present section we investigate lensing features of the obtained wormhole solutions. To this aim we need to study the behavior of null geodesics traveling within the wormhole spacetime. The starting point of our study is the following Lagrangian {describing the motion of photons in equatorial plane $\theta=\pi/2$,}
\be\label{Lagran}
2{\mathfrak L}=g_{\mu\nu}\dot{x}^\mu\dot{x}^\nu=-\dot{t}^2+\left(1-\f{b(r)}{r}\right)^{-1}\dot{r}^2+r^2\dot{\phi}^2,
\ee
where use has been made of the spacetime metric (\ref{evw}) and an overdot denotes derivative with respect to the curve parameter $\eta$. Because of spherical symmetry{, the same results can be applied to all values of $\theta$}. The Lagrangian ${\mathfrak L}(\dot{x},x)$ is constant along a geodesic curve, so one can classify the spacetime geodesics as, timelike geodesics (the world lines of freely falling particles) for which ${\mathfrak L}<0$, lightlike ones for which ${\mathfrak L}=0$ and spacelike geodesics for which ${\mathfrak L}>0$. {Then}, equation of photon trajectory takes {the following} form 
\be\label{effequation}
\dot{r}^2+\left(1-\f{b(r)}{r}\right)\left(\f{h^2}{r^2}-{\mathcal E}^2\right)=0,
\ee
where ${\mathcal E}=\dot{t}$ is the total energy of the particle moving on its orbit {and} $h=r^2\dot{\phi}$ is its specific angular momentum. Consider now a light ray incoming from infinity, reaching the minimum distance $r_{\rm min}$ from the center of the gravitating body, emerging then in another direction. The deflection angle of the light ray as a function of the closet distance approach is given by~\cite{SWeinbergbook}
\be\label{defangleequa}
\Theta(r_{\rm min})=-\pi+2\int_{r_{\rm min}}^{\infty}\f{\mu dr}{\left[\left(r^2-rb(r)\right)\left(r^2-J^2\right)\right]^{\f{1}{2}}},
\ee
where $J=h/{\mathcal E}$ is the impact parameter and $dr/d\phi=0$ at $r=r_{\rm min}$, so we have $J=r_{\rm min}$. {Substituting solution (\ref{bfinalsol}) into the above expression}, we obtain the deflection angle as
\be\label{defangleb}
\Theta(r_{\rm min})=-\pi+2r_{\rm min}\int_{r_{\rm min}}^{\infty}\left(r^2-r_{\rm min}^2\right)^{-\f{1}{2}}\left[r^2-\alpha_1r^2-\alpha_2r^4-\alpha_3r^{2(n+2)}-\alpha_4r^{\f{4-3n^2}{3n+4}}\right]^{-\f{1}{2}}dr.
\ee
In the right panel of Fig. (\ref{fig5}) we have plotted {the} deflection angle as a function closest distance approach using numerical methods. It is therefore seen that the more the closest distance decreases, the more the deflection angle grows. Decreasing $r_{\rm min}$ further causes the light ray to infinitesimally come closer to the photon orbit making it to wind up for a large number of times before emerging out. The deflection angle will diverge, eventually, at a critical value of closet distance approach, $r_{\rm min}^{\rm cr}$ where light ray will loop around a circular photon orbit indefinitely. The set of these orbits constructs the photon sphere satisfying $\dot{r}=\ddot{r}=0$~\cite{Hasse-Perlick}~\cite{Perlicklvr}. In~\cite{Shaikh-novel}, it has been shown that the wormhole throat can act as an effective photon sphere located at $r_{\rm min}^{\rm cr}=r_0$. Hence, as $r_{\rm min}\rightarrow r_0$, the deflection angle increases and diverges at the wormhole throat where an unstable photon sphere is present. As a result, the wormhole can produce infinite number of relativistic images of an appropriately placed light source. This infinite sequence corresponds to infinitely many light rays whose limit curve asymptotically spirals towards the unstable photon sphere~\cite{Hasse-Perlick}. Since the photon sphere coincides with the wormhole throat, such a sphere can be detected utilizing thoroughly and carefully designed modern instruments~\cite{Perlicklvr},\cite{highsensinstru}, providing thus, possible observational proofs for the existence of the wormhole. The deflection angle decreases as $r_{{\rm min}}$ increases beyond $r_0$ (the light bears lesser bending) until the closest distance approach reaches a critical value at which $\Theta(r_{{\rm min}}^\star)=0$, i.e., no deflection of light occurs, see the inset of Fig. (\ref{fig5}). In such a situation, any incoming light ray from infinity that reaches the coordinate distance $r=r_{{\rm min}}^\star$, is scattered back to infinity without any spinning around the wormhole lens. Therefore the light ray does not undergo any net deflection by the gravitating object. For $r_{{\rm min}}>r_{{\rm min}}^\star$ the deflection angle is found to be negative, which can be interpreted as there is a repulsion of light by the wormhole configuration. A negative value for deflection angle has also been reported in gravitational lensing by a naked singularity~\cite{nakedndefan}. {It is noteworthy that,} beside the lensing effects that can play a major role in detecting wormhole configurations, various observational aspects have been {investigated} so far with the aim of probing wormholes {that might be living} in our Universe. {Among these studies} we can quote: the study of particle trajectory in the wormhole spacetime~\cite{parttrajworm}, accretion disks around wormholes~\cite{accrdiskworm} and their gravitational wave signatures~\cite{Gwaveworm}. Another interesting candidate for extracting physical information from the wormhole spacetime is the shadow cast by it or its apparent shape~\cite{wormshadow}. As the shape of the shadow is merely determined by the background metric, the observation of such a phenomenon can provide useful information on the nature of the compact object and under some conditions, this interesting event can provide observational testbed for distinguishing the wormhole from other compact bodies. This idea has motivated many researchers to investigate {observational aspects of wormholes, e.g.,} shadows cast by rotating~\cite{raotwormshadow} and charged wormholes~\cite{cohargeshadow} has been studied and in~\cite{OhgamiSakai2005}, the authors have developed the shadow-like images of wormholes surrounded by optically thin dust. The existence of unstable photon orbits is of crucial importance in studying wormhole shadows as these orbits define the boundary between capture and non-capture of the light rays around a wormhole configuration. Thus, the boundary of the shadow is only determined by the metric of spacetime since it corresponds to the apparent shape of the photon sphere as seen by a distant observer~\cite{raotwormshadow,Falcke2000} (see also~\cite{GRGSHADOWGL} for more details). As the lensing effects and shadows are of significant importance for detecting astrophysical compact objects, especially wormholes, it is about time to probe the existence of such objects utilizing advanced instruments, e.g., the US-led Event Horizon Telescope (EHT) project\footnote{Project website: \url{www.eventhorizontelescope.org}} and the European Black Hole Cam (BHC) project\footnote{Project website: \url{www.blackholecam.org}}.
\section{Concluding Remarks}\label{concluding}
In the present work, we tried to build and study static spherically symmetric wormhole configurations in a subclass of {\sf PGT} Lagrangians that allow for spin-$0^+$ propagating modes. We therefore considered the field equations of {\sf PGT} in the absence of spin of matter and followed the strategy of solving the field equations performed in~\cite{mt}. {It was found that in addition to {\sf GR} terms, new geometric terms that come from spacetime torsion contribute to the {\sf EMT} components}. We then proceeded to find exact spacetimes that admit suitable wormhole configurations, considering the field equation derived from variation of gravitational Lagrangian with respect to the connection i.e., Eq.~(\ref{FESFinal1}). {The exact solution for a constant redshift shows that the wormhole's shape has functional dependence on the torsion component $B(r)$.} Assuming then a power-law behavior for torsion component, $B(r)=B_0r^n$, we provided the allowed regions for three of the model parameters ($n$, $b_6$ and $\mu$) for which the conditions on physical reasonability of the model are respected. {We further realized} that, for the obtained solutions, the supporting matter for wormhole geometry obeys the {\sf WEC} and {\sf NEC}. In order to study the nature of this type of matter, we assumed that its radial and tangential pressure profiles depend linearly on energy density with $r$-dependent state parameters, see Eq. (\ref{eoss}). The behavior of these parameters was investigated and it was found that the matter threading the wormhole configuration can assume different equations of state (at the throat) depending on the model parameters, see Fig. (\ref{fig4}). Furthermore, we provided the allowed values of state parameters at the throat subject to fulfillment of conditions on physical validity of the solutions. The anisotropy parameter for wormhole geometry was studied and it was observed that this parameter could admit different maxima near the throat depending on the growth rate of spacetime torsion. Finally we investigated gravitational lensing effects on the wormhole's surrounding environment and it was found that the deflection angle of the incoming beam of light admits positive, zero and negative values. The situation of vanishing deflection angle occurs at different coordinate radii and the value of each radius depends on the behavior of spacetime torsion near the wormhole throat. We therefore observed that, depending on model parameters, the wormhole configuration can act as a converging or diverging lens. It is worth mentioning that, solutions comprising rich information on wormhole configurations in {\sf PGT} may be found by taking, $i)$ a general form of the red-shift function, $ii)$ a non-zero spin density of matter distribution, and $iii)$ different functionalities of the torsion components. Specially the second case could provide a setting based on which the effects of spin on the geometry of a wormhole configuration can be surveyed. More interestingly, these effects can be helpful in probing the geometrical feature of the spacetime that couples to spin of matter, i.e., the spacetime torsion, via highly sensitive observational instruments. However, regarding these cases, the resultant field equations are too complicated to be solved analytically and more advanced mathematical techniques are needed in order to overcome the problem. Work along this line is currently in progress and the results will be reported as an independent work. 

\end{document}